\documentclass[aps,prb,superscriptaddress,reprint]{revtex4-2}
\usepackage[dvipdfmx]{graphicx}
\usepackage{color}
\usepackage{amsmath}
\usepackage{amsthm}
\usepackage{amssymb}
\usepackage{physics, ulem}
\usepackage{cases}
\usepackage{mathrsfs}
\usepackage{subfigure}
\usepackage{comment}
\usepackage{orcidlink}

\usepackage{hyperref}
\hypersetup{
	citecolor = blue,
	colorlinks = true,
	urlcolor = blue
}
\usepackage{physics2}
\usephysicsmodule{ab}

\newcommand{\lb}{\left(}
\newcommand{\rb}{\right)}
\newcommand{\lB}{\left\{}
\newcommand{\rB}{\right\}}
\newcommand{\LB}{\left[}
\newcommand{\RB}{\right]}

\newcommand{\up}{\uparrow}
\newcommand{\down}{\downarrow}

\newcommand{\vv}[1]{\boldsymbol{#1}}

\newcommand{\into}[2]{\left. {#1} \right|_{#2}}

\usepackage{scalefnt} 
\usepackage{mathrsfs} 
\newcommand{\cR}{{{\mbox{\scalefont{0.97}$\mathcal{R}$}}}}
\newcommand{\scR}{{{\mbox{\scalefont{0.7}$\mathcal{R}$}}}} 
\newcommand{\sscR}{{\scalefont{0.5}\cR}}
\newcommand{\cL}{{{\mbox{\scalefont{0.97}$\mathcal{L}$}}}}
\newcommand{\scL}{{{\mbox{\scalefont{0.7}$\mathcal{L}$}}}} 
\newcommand{\sscL}{{\scalefont{0.5}\cL}}
\newcommand{\cN}{{{\mbox{\scalefont{0.95}$\mathcal{N}$}}}}

\newcommand{\sscN}{{{\mbox{\scalefont{0.2}$\mathcal{N}$}}}}

\definecolor{darkgreen}{rgb}{0,0.5,0}

\begin{document}

\title{AC/DC spin current in ferromagnet/superconductor/normal metal trilayer systems}

\author{Koki Mizuno\, \orcidlink{0009-0004-8276-1045}}
\affiliation{Department of Physics, Nagoya University, Nagoya 464-8602, Japan}
\author{Hirone Ishida\, \orcidlink{0009-0000-7095-7105}}
\affiliation{Department of Physics, Saitama University, Saitama 338-8570, Japan}
\author{Manato Teranishi}
\affiliation{Department of Physics, Nagoya University, Nagoya 464-8602, Japan}

\begin{abstract}
	Spin pumping with superconductors has been extensively studied, particularly in double-layer systems. 
	In this study, we investigate spin pumping in a trilayer system comprising a ferromagnetic insulator (FMI), a superconductor (SC), and a normal metal (NM). 
	We derive the AC and DC spin currents in the NM layer induced by spin motion in the FMI under circularly polarized microwave irradiation. 
	If we treat the spin motion as classical, the AC spin current is expressed.
	On the other hand, if we treat the spin motion as quantum quasiparticles, the DC spin current is derived.
	After these derivations, while the computational cost of evaluating the spin current is extremely high, we mitigate this using the Quantics Tensor Cross Interpolation (QTCI) method. 
	We present numerical results showing the dependence of the spin current on temperature, microwave frequency, and superconductor layer thickness. 
	Notably, the temperature dependence of AC and DC spin currents exhibits a coherence peak. 
	Furthermore, we have discovered a transition structure in the dependence of the spin current on the thickness of the superconductor layer, where the dependence changes after a particular frequency.
\end{abstract}

\date{\today}
\maketitle

\section{Introduction}
Spin transport phenomena in hybrid structures involving superconductors (SCs) have emerged as a central topic in modern condensed matter physics, motivated by both fundamental interest and potential applications in low-dissipation spintronic devices \cite{linder2015superconducting_nature, spin-polarized-tunneling-1994, PRL-1970-splitting, RevModPhys-spintronics-2004,beckmann2016spin, spin-pumping-2002-FMINM, Ohnuma-dc-spin-pumping, HIROHATA2020166711,cai2023superconductor}.
In particular, the interplay between spin dynamics and superconductivity is known to produce rich and unconventional behaviors due to the coexistence of spin-singlet pairing and nonequilibrium spin currents.
Among the mechanisms of spin transport, spin pumping—wherein a dynamic magnetization injects spin angular momentum into an adjacent conductor—has been extensively studied in bilayer structures consisting of a ferromagnetic insulator (FMI) and a superconductor \cite{Kato-spin-SC-swave, Ominato-FMR-dwave, Ominato-FMR-pwave}.
These studies have revealed key insights into spin relaxation \cite{Spin-relaxation-mesoscopic-SC-2008,SC-spin-transport-relaxation-2002,Nuclear-Spin-Relaxation-Al}, spin mixing conductance \cite{han2020spin,spin-mixing-PRB-2017}, and coherence effects \cite{iop-escrig2015spin, umeda2018spin,Sun-spin-pumping-2023} in the superconducting state.

More recently, attention has shifted toward trilayer systems of the form FMI/SC/normal metal (NM), which are of particular interest for their potential to generate and detect spin currents via the inverse spin Hall effect (ISHE) \cite{Jen-trilayer-2018, Muller-trilayer-2021, Rogdakis-trilayer-2019, suraj-trilayer-2020}. 
Experimental reports on such trilayers have demonstrated signatures of spin pumping through the SC layer, even below the superconducting transition temperature. 
However, the theoretical understanding of such structures remains incomplete. 
In particular, while bilayer models have been extensively explored, there exists a clear lack of microscopic theoretical treatments addressing the generation and transmission of AC and DC spin currents in trilayer FMI/SC/NM geometries.

In this work, we address this gap by developing a comprehensive microscopic theory of spin pumping in FMI/SC/NM trilayer systems. 
We derive the AC and DC spin currents generated in the NM layer as a result of magnetization dynamics in the FMI under circularly polarized microwave irradiation. 
The derivation is carried out within the Keldysh Green’s function formalism \cite{kamenev2023field}. 
The AC spin current is obtained by treating the FMI spin dynamics classically, while the DC spin current arises from a quantum treatment with magnons.

To perform the required numerical integration in momentum-frequency space, we employ the Quantics Tensor Cross Interpolation (QTCI) \cite{ritter2024quantics, ishida2024low, rohshap2024two,sroda2024high}.
This method discretizes the integrand using the quantics representation \cite{oseledets2009approximation, gourianov2022quantum, shinaoka2023multiscale} and constructs a low-rank Quantics Tensor Train (QTT) via Tensor Cross Interpolation (TCI)\cite{dolgov2020parallel, nunez2022learning, erpenbeck2023tensor, fernandez2024learning}.
This enables exponentially accurate integration with high speed and low memory cost, allowing for detailed theoretical and numerical investigation of spin transport in the trilayer system.

Our results reveal that both AC and DC spin currents exhibit coherence peaks near the superconducting critical temperature, and display a crossover in thickness dependence, transitioning from exponential decay to non-exponential behavior at certain microwave frequencies. 
These features point to interference effects of quasiparticles in the SC layer and may serve as theoretical signatures for future experimental verification.

To our knowledge, this is the first study to present a unified microscopic derivation of both AC and DC spin currents in FMI/SC/NM trilayers, and the first to incorporate QTCI in the context of spin transport. 
Our work provides a solid foundation for understanding spin pumping through superconductors in realistic geometries and opens the door to further studies in superconducting spintronics.

Our paper is organized as follows.
First, we formulate the spin current injected into the NM layer of the FMI/SC/NM system in Sec.~\ref{sec-formulation}.
In Sec.~\ref{sec-numerical}, we introduce the numerical method for calculating the spin current.
In Sec.~\ref{sec-result}, we show the numerical results for the dependence of the spin current defined in the NM layer on the temperature, microwave frequency, and the SC layer thickness.
Finally, we conclude our paper in Sec.~\ref{sec-conclusion}.




\section{Formulation}\label{sec-formulation}

\begin{figure}
    \centering
    \includegraphics[scale=0.9]{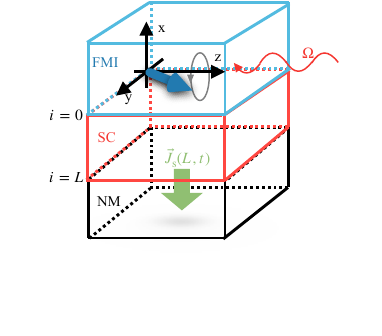}
    \caption{Schematic illustration of the trilayer system considered in this study.
    The spin current, denoted by $\vec{J}_{\rm s}(L,t)$, is injected into the top of the NM layer at time $t$ by a circularly polarized microwave field with frequency $\Omega$.
    The spin-space coordinate axes $(x,y, z)$ are defined such that the $z$-axis is aligned with the direction of microwave propagation.
    The interfaces between the FMI and SC, and between the SC and NM, are located at $z = 0$ and $ z = L$, respectively.
    }
    \label{fig-sys}
\end{figure}

We consider the trilayer system illustrated in FIG.~\ref{fig-sys}.
This section presents the formulation of the spin current injected into the top of the NM layer through the SC layer under the microwave excitation of the FMI layer.

\subsection{Model construction for trilayer system}
We begin by defining the Hamiltonian of the system, which is given by
\begin{align}
    H &= H_{\rm FMI} + H_{\rm S-h}(t) + H_{\rm SC} + H_{\rm NM} + H_{\rm int}, 
    \\
    H_{\rm FMI} &= -J \sum_{\langle i,j \rangle \in \Lambda_{\rm H}} \vv{S}_i \cdot \vv{S}_j -\gamma h_{z}\sum_{i \in \Lambda_{\rm H}} S_{i,z} 
    \label{Heisenberg}
    \\
    H_{\rm S-h}(t)
    &= -\gamma \sum_{i \in \Lambda_{\rm H}}\LB h_{+}(t) S_{i,-} + h_{-}(t) S_{i,+} \RB,
    \label{int_spin-field}
    \\
    H_{\rm SC} &= \sum_{i,i'\in \Lambda_{\rm SC}}\sum_{\sigma,\sigma'}\sum_{\vv{k}\in \rm BZ} \hat{d}_{i\sigma,\vv{k}}^{\dag} \mathcal{H}^{i\sigma,i'\sigma'}_{\rm SC}(\vv{k}) \hat{d}_{i'\sigma',\vv{k}},
    \\
    H_{\rm NM} &= \sum_{i,i' \in \Lambda_{\rm NM}}\sum_{\sigma,\sigma'} \sum_{\vv{k} \in \rm BZ} \hat{c}_{i\sigma,\vv{k}}^{\dag} \mathcal{H}^{i\sigma,i'\sigma'}_{\rm NM}(\vv{k}) \hat{c}_{i'\sigma',\vv{k}},
    \\
    H_{\rm int} &= \sum_{\sigma\vv{k}} \left( 
        T_{\rm int} \hat{d}_{L-1\sigma,\vv{k}}^{\dag} \hat{c}_{L\sigma,\vv{k}}  
        \right. 
        \notag\\ 
        &\left.
        +
        \sum_{\vv{q}}V S_{+,\vv{q}} \hat{d}_{0\sigma',\vv{k}+\vv{q}}^{\dag} \sigma_{-}^{\sigma'\sigma}\hat{d}_{0\sigma,\vv{k}} + \text{h.c.} 
        \right),
\end{align}
where the Hamiltonian $H_{\rm FMI}$ is the Heisenberg model defined on the two-dimensional lattice $\Lambda_{\rm H}$, and the $\langle i,j \rangle$ denotes the sums of nearest neighbor sites.
The spin operator $\vv{S}_i$ is defined as $\vv{S}_i = (S_{i,x}, S_{i,y}, S_{i,z})$, where $S_{i,\alpha}$ is the spin operator at site $i \in \Lambda_{\rm H}$ in the $\alpha$-direction.
The coefficient $J$ is the exchange interaction between the nearest neighbor spins.
We consider an interaction between the Heisenberg spin and the external microwave field written as $-\gamma\sum_{i}\vv{h}(t)\cdot\vv{S}_{i}$, where $\gamma$ is the gyromagnetic ratio and $\vv{h}(t) = (h_{x}(t), h_{y}(t), h_{z})$ is the external magnetic field.
In this Hamiltonian, we define $S_{i,\pm} = S_{i,x} \pm i S_{i,y}$ and $h_{\pm}(t) = h_{x}(t) \pm i h_{y}(t)$.
Here, we assume that the external field is circularly polarized in the $x$-$y$ plane, i.e., $h_{\pm}(t) = he^{\mp i\Omega}$ and $\partial_{t}h_{z}=0$, where $\Omega$ is the microwave frequency and $h$ is the amplitude of the microwave field.
The static part of the interaction ($h_z$) and the microwave part ($h_{\pm}(t)$) are included into $H_{\rm FMI}$ and $H_{\rm S-h}(t)$, respectively.
The Hamiltonians $H_{\rm SC}$ and $H_{\rm NM}$ are Hamiltonians of a superconducting and normal metal layer defined on the lattice $\Lambda_{\rm SC}$ and $\Lambda_{\rm NM}$, where these lattices defined as $\Lambda_{\rm SC} = \lB i  \mid 0 \leq i \leq L-1,\, i \in \mathbb{Z} \rB$ and $\Lambda_{\rm NM} = \lB i \mid L \leq i, \, i \in \mathbb{Z}  \rB$, i.e., the superconducting and normal metal layers are defined on finite and half-infinite lattice, respectively.
The operator $\hat{d}_{i\sigma,\vv{k}}^{\dag}$ ($\hat{d}_{i\sigma,\vv{k}}$) and $\hat{c}_{i\sigma,\vv{k}}^{\dag}$ ($\hat{c}_{i\sigma,\vv{k}}$) are the creation (annihilation) operator of the electron with spin $\sigma \in \{ \up, \down \}$ and in-plane momentum $\vv{k} \in \rm BZ$ in the SC and NM layer, respectively, where BZ denotes the first Brillouin zone in the two-dimensional momentum space perpendicular to the interface.
We consider the interaction between the SC and NM layer with amplitude $T_{\rm int}$, and the FMI and SC layer with amplitude $V$ defined in $H_{\rm int}$. 
In the second term of $H_{\rm int}$, we define the matrix $\sigma_{\pm} = (\sigma_{x} \pm i \sigma_y)/2$.
The operator $S_{+,\vv{q}}$ is the Fourier component of the spin operator defined as $S_{+,\vv{q}} = (1/\sqrt{N})\sum_{i\in \Lambda_{\rm H}} S_{i,+} e^{-i\vv{q}\cdot\vv{r}_i}$, where $\vv{r}_i$ is the position of the site $i$ and $N$ is the number of sites in the Heisenberg model.

In this paper, we consider the $s$-wave superconducting layer, thus the Hamiltonian $\mathcal{H}^{i\sigma,i'\sigma'}_{\rm SC}(\vv{k})$ is given by
\begin{align}
    H_{\rm SC} &= \sum_{\sigma,\sigma'}\sum_{i,i' \Lambda_{\rm SC}}\sum_{\vv{k}} \bar{d}_{i\sigma,\vv{k}} \mathcal{H}^{i\sigma,i'\sigma'}_{\rm SC}(\vv{k}) d_{i'\sigma',\vv{k}},
    \\
    \mathcal{H}_{\rm SC}(\vv{k})|_{i' = i}
    &= \mqty(
        \mathcal{H}_{0}(\vv{k})   &   \Delta_{T} \\
        \Delta_{T}^{\dag}    &  -\mathcal{H}_{0}(-\vv{k})
    ),
    \notag\\ 
    \mathcal{H}_{0}(\vv{k}) &= \LB -2t(\cos k_x + \cos k_y) - \mu \RB \sigma_{0},
    \notag\\ 
    \mathcal{H}_{\rm SC}(\vv{k})|_{i' = i+1}
    &= \mqty(
       -t\sigma_{0}  &  O_{2} \\ 
       O_{2}        &  t\sigma_{0}
    ),
\end{align}
where the pair potential in the onsite term is defined at temperature $T < T_{\rm C}$ as 
\begin{align}
    \Delta_{T} &= i\Delta_{0}(T) \sigma_{y},
    \notag\\
    \Delta_{0}(T) &= 1.76 k_{\rm B} T_{\rm C}\tanh\lb 1.74\sqrt{\frac{T_{\rm C} - T}{T}} \rb, 
    \label{def-pair-potential}
\end{align}
where $T_{\rm C}$ is the critical temperature, and the coefficient of the right-hand side in terms of $\Delta_{0}(T)$ is phenomenologically determined by the BCS theory.
If the temperature is $T > T_{\rm C}$, the pair potential is set $\Delta_{0}(T) = 0$.

\subsection{Spin current in the NM layer}
The spin current on the position $x$ at time $t$ is defined as
\begin{align}
    \vec{J}_{\rm s}(x,t) 
    &= \expval{\partial_{t} \widehat{\vv{S}}(x,t)}
    =
    -i\expval{\comm{\widehat{\vv{S}}(x,t)}{H}},
    \notag\\
    \widehat{\vv{S}}(x,t)
    &= \sum_{\sigma,\sigma'} \sum_{\vv{k} \in \rm BZ}  \hat{c}_{x\sigma',\vv{k}}^{\dag}(t) \vec{{\sigma}}^{\sigma'\sigma} \hat{c}_{x\sigma,\vv{k}}(t),
    \label{def-spin-current}
\end{align}
where $\vec{\sigma} := (\sigma_{x}, \sigma_{y}, \sigma_{z})$ is the vector of Pauli matrix.
The expectation value $\expval{\hat{O}}$ is defined as
\begin{align}
    \expval{\hat{O}} &= \frac{\int \prod_{\psi}\mathcal{D}[\bar{\psi}\psi] 
    \hat{O} e^{ iS}}
    {\int \prod_{\psi}\mathcal{D}[\bar{\psi}\psi] e^{iS}},
\end{align}
where the $\psi$ is arbitrary field in the system, and $S$ is the action of the system.
If we define the spin operator $\widehat{\vv{S}}(x,t)$ as the spin operator at the top of the NM layer, i.e., $x = L$, we can find that the injected spin current into the normal metal is written as 
\begin{align}
    \vec{J}_{\rm s}(L,t) =& \frac{T_{\rm int}}{2}\sum_{\sigma,\sigma'}\sum_{\vv{k} \in \rm BZ}
    \Im \LB \expval{ c_{L\sigma',\vv{k}}^{\dag}(t)  \vec{\tilde{\sigma}}^{\sigma'\sigma} d_{L-1\sigma,\vv{k}}(t)} \RB,
    \notag\\ 
    \vec{\tilde{\sigma}}
    = &
    \mqty(
        \vec{\sigma} & 0 \\
        0            & ^{t}\vec{\sigma}
    ).
    \label{J-spin-ope}
\end{align}
A vector with arrows denotes the spin polarization direction of the spin current. 
In the following sections, we derive the concrete expression of the spin current in the NM layer by perturbation expansions of the interaction $S_{\rm int}$.
Depending on the order of the perturbation, we can obtain the AC and DC spin currents, which correspond to the second and third-order perturbations, respectively.

\subsection{AC spin current}
First, we derive the AC spin current.
Here, we treat the spin in the FMI layer as a classical value.
Thus, the expectation value in Eq.~(\ref{J-spin-ope}) is calculated with action $S$ defined as
\begin{align}
    S &= S_{\rm SC} + S_{\rm NM} + S_{\rm int}^{\rm cl}, \label{S-AC}
    \\
    S_{\rm SC} &
    = \sum_{i,i'\in \Lambda_{\rm SC}} \sum_{\sigma,\sigma'}\sum_{\vv{k} \in \rm BZ}\int dt
    \notag\\ 
    &\times\LB \bar{d}_{i\sigma,\vv{k}} \lb i\partial_t - \mathcal{H}^{i\sigma,i'\sigma'}_{\rm SC}(\vv{k}) \rb d_{i'\sigma',\vv{k}} \RB,
    \\ 
    S_{\rm NM} &
    = \sum_{i,i'\in \Lambda_{\rm NM}} \sum_{\sigma,\sigma'}\sum_{\vv{k} \in \rm BZ}\int dt 
    \notag\\ 
    &\times
    \LB \bar{c}_{i\sigma,\vv{k}} \lb i\partial_t - \mathcal{H}^{i\sigma,i'\sigma'}_{\rm NM}(\vv{k}) \rb c_{i'\sigma',\vv{k}} \RB,
    \\
    S_{\rm int}^{\rm cl} &= -\sum_{\sigma\sigma'\vv{k}} \left( 
        T_{\rm int} \hat{d}_{L-1\sigma,\vv{k}}^{\dag} \tau_{z} \hat{c}_{L\sigma,\vv{k}}  
        \right. 
        \notag\\ 
        &\left.
        +
        V S_{+}(t) \hat{d}_{0\sigma',\vv{k}}^{\dag} \bar{\sigma}_{-}^{\sigma'\sigma}\hat{d}_{0\sigma,\vv{k}} + \text{h.c.} 
        \right),
\end{align}
where we use the Grassmann number as fermionic field $\bar{d}_{x\sigma,\vv{k}}$ and $d_{x\sigma,\vv{k}}$ (Nambu spinor) for the SC, $\bar{c}_{x\sigma,\vv{k}}$ and $c_{x\sigma,\vv{k}}$ is electrons in the NM, and $\bar{\sigma}_{\pm} := \sigma_{\pm} \oplus (-\sigma_{\mp})$.
We define the matrix $\tau_z$ as the $z$ component of the Pauli matrix with the particle-hole basis.
The $S_{\pm}(t)$ is the classical value of the spin operator in the FMI layer written as $S_{\pm}(t) = S e^{\mp i\Omega t}$ for the case of a circularly polarized microwave field.
According to the second-order perturbations of $S_{\rm int}^{\rm cl}$, we can derive the AC spin current as
\begin{widetext}
    \begin{align}
        \vec{J}_{\rm s}^{\rm AC}(L, \Omega)
        &= -\frac{T_{\rm int}^2 V S}{8}\sum_{\alpha=\pm} \sum_{\sigma_1,\sigma_2,\sigma_2'}\sum_{\sigma,\sigma'} \sum_{\vv{k} \in \rm BZ}
        \int d\omega \vec{\tilde{\sigma}}^{\sigma'\sigma}
        \notag\\
        &\times
        \Re
        \left[
             G_{\rm SC}^{R,L-1 \sigma, 0\sigma_2}(\vv{k},\omega+\alpha\Omega)
            \bar{\sigma}_{-\alpha}^{\sigma_2 \sigma_2'}
            G_{\rm SC}^{R,0\sigma_2', L-1 \sigma_1}(\vv{k},\omega)
            \tau_{z} 
            G_{\rm NM}^{<.L\sigma_1, L\sigma'}(\vv{k},\omega)
        \right.
        \notag\\ 
        &\quad\quad+
            G_{\rm SC}^{R,L-1 \sigma, 0\sigma_2}(\vv{k},\omega+\alpha\Omega)
            \bar{\sigma}_{-\alpha}^{\sigma_2 \sigma_2'}
            G_{\rm SC}^{<,0\sigma_2', L-1 \sigma_1}(\vv{k},\omega)
            \tau_{z} 
            G_{\rm NM}^{A.L\sigma_1, L\sigma'}(\vv{k},\omega)
        \notag\\ 
        &\left.\quad\quad+
            G_{\rm SC}^{<,L-1 \sigma, 0\sigma_2}(\vv{k},\omega+\alpha\Omega)
            \bar{\sigma}_{-\alpha}^{\sigma_2 \sigma_2'}
            G_{\rm SC}^{A,0\sigma_2', L-1 \sigma_1}(\vv{k},\omega)
            \tau_{z} 
            G_{\rm NM}^{A.L\sigma_1, L\sigma'}(\vv{k},\omega)
        \right],
        \label{AC-spin-current}
    \end{align}
\end{widetext}
where $G_{\rm NM}^{A,R,<,L,L}(\vv{k},\epsilon)$ and $G_{\rm SC}^{A,R,<,L/0,0/L}(\vv{k},\epsilon)$ are the Fourier component of the Green's function defined with contour ordered product $T_{\mathcal C}$ as 
\begin{align}
    G_{\rm NM}(\vv{k},t_1,t_2) &:= -i \expval{T_{\mathcal{C}} c_{L\sigma_{1},\vv{k}}(t_1) c_{L\sigma',\vv{k}}^{\dag}(t_2) }_{0},
    \notag\\
    G_{\rm SC}(\vv{k},t_1,t_2) &:= -i \expval{T_{\mathcal{C}} d_{L-1/0\sigma_{1},\vv{k}}(t_1) d_{0/L-1\sigma',\vv{k}}^{\dag}(t_2) }_{0},
    \label{Green-SC-NM}
\end{align}
where the expectation value $\expval{\hat{O}}_{0}$ is calculated with the action without the interaction $S_{\rm int}^{\rm cl}$.
The Green's functions $G_{\rm NM}^{A, R,<, L, L}(\vv{k},\epsilon)$ and $G_{\rm SC}^{A, R,<, L/0,0/L}(\vv{k},\epsilon)$ are the advanced, retarded, and lesser part of them.
Details of this derivation are given in the appendix \ref{app-AC}.

\subsection{DC spin current}
This section treats the spin motion in the FMI layer as a quantum field to derive the DC spin current.
Thus, we define the creation and annihilation operator of magnon from the Holstein-Primakov transformation $S_{i,+} =\sqrt{2s} a_{i}$, $S_{i,-} = \sqrt{2s} a_{i}^{\dag}$ and $S_{i, Z} = s - a_{i}^{\dag}a_{i}$, and its Fourier transformation $a_{i} = (1/\sqrt{N})\sum_{\vv{q}\in \rm BZ} a_{\vv{q}}e^{-i\vv{q}\cdot\vv{r}_i}$, where $\vv{r}_i$ is the position of the site $i$ and $s$ is the spin quantum number \cite{Holstein_Primakoff}.
Therefore, the Hamiltonians (\ref{Heisenberg}) and (\ref{int_spin-field}) are written as
\begin{align}
    H_{\rm M} &= \sum_{\vv{q}} \omega_{\vv{q}} a_{\vv{q}}^{\dag}a_{\vv{q}},
    \quad 
    H_{\rm M-h}(t) = \lambda ( h_{+}(t) a_{\vv{0}}^{\dag} + h_{-}(t) a_{\vv{0}} ), 
    \notag\\ 
    \omega_{\vv{q}} &= 2sJ \LB n_{\rm d} - 2\sum_{<d_{ij}>}\cos \vv{d}_{ij}\cdot \vv{q} \RB + \gamma h_z,
    \label{H-Mag}
\end{align}
where $\lambda = -\gamma \sqrt{2sN}$.
The c-numbers $\bar{a}_{\vv{q}}$ and $a_{\vv{q}}$ are bosonic field for the magnon.
The integer $n_{\rm d}$ is the number of nearest neighbor sites, $\vv{d}_{ij}$ is the vector connecting the nearest-neighbor sites $i$ and $j$.
The sum $\sum_{<d_{ij}>}$ denotes the summation over all nearest-neighbor pairs.
Therefore, we calculate the expectation value of the spin current in the NM layer based on the action
\begin{align}
    S &= S_{\rm M} + S_{\rm M-h} + S_{\rm SC} + S_{\rm NM} + S_{\rm int},
    \\
    S_{\rm M} &= \sum_{\vv{q}}\int dt \LB \bar{a}_{\vv{q}} \lb i\partial_t - \omega_{\vv{q}} \rb a_{\vv{q}} \RB,
    \\
    S_{\rm M-h} &= -\int dt \, H_{\rm M-h} (t),
    \\ 
    S_{\rm int} &= -\sum_{\sigma\vv{k}\vv{q}}\int dt \left[
        T_{\rm int}\bar{d}_{L-1\sigma,\vv{k}} \tau_{z} c_{L\sigma,\vv{k}}
    \right.
        \notag \\
        &\left. \quad\quad\quad\quad
        +
        Va_{\vv{q}}  \bar{d}_{0\sigma',\vv{k}+\vv{q}}\bar{\sigma}_{-}^{\sigma'\sigma} d_{0\sigma,\vv{k}} + \text{h.c.}
        \right].\label{int-trilayer}
\end{align}

The goal of the derivation of the DC spin current is to define the pumped spin current as
\begin{align}
    \Delta \vec{J}_{\rm s}^{(3)}(L,\omega_{\rm s}) := \vec{J}_{\rm s}^{(3)}(L,\omega_{\rm s}) - \into{\vec{J}_{\rm s}^{(3)}}{h=0}(L,\omega_{\rm s}),
\end{align}
where $\vec{J}_{\rm s}^{(3)}(L,\omega_{\rm s})$ is the Fourier component of the spin current with third-order perturbation expansion of $S_{\rm int}$ in the presence of the external field.
In this opinion, the expansions must include the Green's function of the magnon $G_{\rm M}(\vv{q},\omega)$ defined as
\begin{align}
    G_{\rm M}(\vv{q},t_1,t_2) &=-i \expval{\mathcal{T}_{\mathcal C} a_{\vv{q}}(t_1) a_{\vv{q}}^{\dag}(t_2)}_{\rm M},
    \notag\\ 
    \expval{\hat{O}}_{\rm M} &:= \frac{\int \mathcal{D}[\bar{a}a]  
    \hat{O} e^{ iS_{\rm M} + iS_{\rm M-h}} }{\int \mathcal{D}[\bar{a}a] e^{ iS_{\rm M} + iS_{\rm M-h}}} .
\end{align}
Simultaneously, the expanded expectation value must be a quadratic for all field values.
Therefore, the lowest-order terms of this perturbation expansion are the third-order terms.
Since third-order terms include only one magnon $G_{\rm M}(\vv{q},\omega)$, the pumped spin current is written with the correction term $\Delta G_{\rm M}(\vv{q},\omega)$ coming from the effect of ferromagnetic resonance, which is given as
\begin{align}
    \Delta G^{<}_{\rm M}(\vv{q},\omega) = h' \delta_{\vv{q},\vv{0}} \delta(\omega - \Omega) \abs{G_{\rm M}^{(0)\rm R}(\vv{q},\omega)}^2,
\end{align}
where $h' = -2i\pi \lambda^2 h^2$ and $G_{\rm M}^{(0)\rm R}(\vv{q},\omega)$ is the retarded part of the Green's function defined as 
\begin{align}
    G_{\rm M}^{(0)}(\vv{q},t_1,t_2) &=-i \expval{\mathcal{T}_{\mathcal C} a_{\vv{q}}(t_1) a_{\vv{q}}^{\dag}(t_2)}_{0,\rm M},
\end{align}
where this expectation value is calculated with the action without the interaction $S_{\rm M-h}$.
The derivation of this correction is seen in the appendix \ref{app-mag}.
Considering the Langreth rule, we can derive the pumped spin current
\begin{widetext}
\begin{align}
    \Delta \vec{J}_{\rm s}^{(3)}(L,\omega_{\rm s}) 
    =& \frac{ \pi (\lambda h T_{\rm int} V)^{2}}{3} \abs{G_{\rm M}^{(0)\rm R}(\vv{0},\Omega)}^2 \delta(\omega_{\rm s}) \int d\epsilon \sum_{\vv{k} \in \rm BZ} \Re \Tr 
    \notag\\ 
    & \vec{\tilde{\sigma}}
    \left[
        G_{\rm SC}^{R,L-1,0}(\vv{k},\epsilon)  \bar{\sigma}_{-} G_{\rm SC}^{R,0,0}(\vv{k},\epsilon-\Omega) \bar{\sigma}_{+} G_{\rm SC}^{R,0,L-1}(\vv{k},\epsilon) \tau_z  G_{\rm NM}^{<,L,L}(\vv{k},\epsilon) 
    \right.
    \notag\\
    &+ G_{\rm SC}^{R,L-1,0}(\vv{k},\epsilon)  \bar{\sigma}_{-} G_{\rm SC}^{R,0,0}(\vv{k},\epsilon-\Omega) \bar{\sigma}_{+} G_{\rm SC}^{<,0,L-1}(\vv{k},\epsilon)\tau_z G_{\rm NM}^{A,L,L}(\vv{k},\epsilon)
    \notag\\
    &+ G_{\rm SC}^{R,L-1,0}(\vv{k},\epsilon)  \bar{\sigma}_{-} G_{\rm SC}^{<,0,0}(\vv{k},\epsilon-\Omega) \bar{\sigma}_{+} G_{\rm SC}^{A,0,L-1}(\vv{k},\epsilon)\tau_z G_{\rm NM}^{A,L,L}(\vv{k},\epsilon)
    \notag\\
    &+ G_{\rm SC}^{<,L-1,0}(\vv{k},\epsilon)  \bar{\sigma}_{-} G_{\rm SC}^{A,0,0}(\vv{k},\epsilon-\Omega) \bar{\sigma}_{+} G_{\rm SC}^{A,0,L-1}(\vv{k},\epsilon) \tau_z G_{\rm NM}^{A,L,L}(\vv{k},\epsilon)
    \notag\\ 
    &+ G_{\rm SC}^{R,L-1,0}(\vv{k},\epsilon)  \bar{\sigma}_{+} G_{\rm SC}^{R,0,0}(\vv{k},\epsilon+\Omega) \bar{\sigma}_{-} G_{\rm SC}^{R,0,L-1}(\vv{k},\epsilon) \tau_z G_{\rm NM}^{<,L,L}(\vv{k},\epsilon)
    \notag\\
    &+ G_{\rm SC}^{R,L-1,0}(\vv{k},\epsilon)  \bar{\sigma}_{+} G_{\rm SC}^{R,0,0}(\vv{k},\epsilon+\Omega) \bar{\sigma}_{-} G_{\rm SC}^{<,0,L-1}(\vv{k},\epsilon) \tau_z G_{\rm NM}^{A,L,L}(\vv{k},\epsilon)
    \notag\\
    &+ G_{\rm SC}^{R,L-1,0}(\vv{k},\epsilon)  \bar{\sigma}_{+} G_{\rm SC}^{<,0,0}(\vv{k},\epsilon+\Omega) \bar{\sigma}_{-} G_{\rm SC}^{A,0,L-1}(\vv{k},\epsilon) \tau_z G_{\rm NM}^{A,L,L}(\vv{k},\epsilon)
    \notag\\
    &+\left. G_{\rm SC}^{<,L-1,0}(\vv{k},\epsilon)  \bar{\sigma}_{+} G_{\rm SC}^{A,0,0}(\vv{k},\epsilon+\Omega) \bar{\sigma}_{-} G_{\rm SC}^{A,0,L-1}(\vv{k},\epsilon) \tau_z G_{\rm NM}^{A,L,L}(\vv{k},\epsilon)
    \right],
    \label{pumped-spin-current}
\end{align}
\end{widetext}
where the definitions of the Green's functions are the same as Eq.~(\ref{Green-SC-NM}).
From $\delta(\omega_{\rm s})$, these terms correspond to the static values of the pumped spin current with $\omega_{\rm s} = 0$.
Thus, we could get the DC spin current in the NM layer.
As a comment, we can derive the AC spin current by treating the motion of the magnetization in the FMI layer as a given and classical value.
The expectation value $\expval{\hat{O}}_0$ refers to that obtained within the non-interacting theory.
For simplicity, we represent the pumped spin current (\ref{pumped-spin-current}) as 
\begin{align}
    \Delta \vec{J}_{\rm s}^{(3)}(L,\omega_{\rm s}) 
    &= J_{0} \delta(\omega_{s}) \vec{\mathcal{J}}_{\rm s}(L),
    \label{pumped-spin-current-simp}
    \\
    J_{0} &:= \frac{ \pi (\lambda h T_{\rm int} V)^{2}}{3} \abs{G_{\rm M}^{(0)\rm R}(\vv{0},\Omega)}^2.
\end{align}
For a rigorous derivation of this quantity by algebraic methods, see the appendix \ref{app-spin-current}.
To integrate this, we need to calculate the lesser part of the real-space Green's function for the SC layer and the half-infinite NM layer.
These steps are seen in the appendices \ref{app-SC} and \ref{app-NM}.

\subsection{temperature dependence and normalization}
This section considers the temperature dependence of the pumped spin current $\Delta \vec{J}_{\rm s}(L,\omega_{\rm s})$.
In the expression Eq.~(\ref{pumped-spin-current}), the temperature dependency is only in the pair potential of the superconducting layer.
From the definition of pair potential (\ref{def-pair-potential}), the spin current is independent of the temperature at $T > T_{\rm C}$.
Therefore, the AC spin current given by Eq.~(\ref{AC-spin-current}) can be normalized by
\begin{align}
    J^{\rm AC,norm}_{\mathrm{s},i } (L,\Omega)
    &:= \frac{J_{\mathrm{s},i}^{\rm AC}(L,\Omega)}{J_{\mathrm{s},i}^{\rm AC}(L,\Omega)|_{T>T_{\rm C}}}.
    \label{J-AC-norm}
\end{align}

In the same way, we normalize the pumped DC spin current by its value with $T > T_{\rm C}$.
According to this normalization, we represent in a way of the Eq.~(\ref{pumped-spin-current-simp}) as
\begin{align}
    \Delta {J}_{i}^{\rm norm} &:=
    \frac{\Delta {J}_{\mathrm{s},i}^{(3)}(L,\omega_{\rm s})}{\Delta {J}_{\mathrm{s},i}^{(3)}(L,\omega_{\rm s})|_{T>T_{\rm C}}}
    = C^{-1} {\mathcal{J}}_{\mathrm{s},i}(L),
    \notag\\ 
    C &= {\mathcal{J}}_{\mathrm{s},i}(L)|_{T>T_{\rm C}}.
    \label{J_renom-T}
\end{align}

\section{Numerical Method}\label{sec-numerical}


In this section, we describe a numerical integration scheme based on Quantics Tensor Cross Interpolation (QTCI)~\cite{ritter2024quantics, ishida2024low, rohshap2024two}. While the method is equivalent to a Riemann sum, recent studies have demonstrated that it is highly effective for high-dimensional integration problems due to the compression properties of QTCI. 

\subsection{Quantics tensor cross interpolation}
We consider a high-dimensional function $f(\vv{u})$, where $\vv{u} = (u_1, u_2, \cdots, u_{\sscN}) \in [0, 1)^\sscN$ is a vector of continuous variables. To efficiently compress and evaluate such functions, we employ the Quantics Tensor Cross Interpolation (QTCI) method, which consists of two key steps: (1) discretization using the quantics representation~\cite{gourianov2022quantum, shinaoka2023multiscale}, and (2) adaptive low-rank decomposition via Tensor Cross Interpolation (TCI)~\cite{dolgov2020parallel, fernandez2024learning}.

Each variable $u_i$ is mapped onto a binary grid using its $\cR$-bit expansion:
\begin{align}
    u_i = (0.\sigma^{(i)}_1 \sigma^{(i)}_2\cdots\sigma^{(i)}_{\sscR})_2 = \sum_{r=1}^{\sscR}\sigma^{(i)}_{r}2^{-r},
\end{align}
where $\sigma^{(i)}_r \in \{0, 1\}$. Applying this to all $\cN$ variables results in a discretized tensor $F_{\vv{\sigma}}$ with $\cL = \cR\cN$ binary indices:
\begin{align}
    F_{\vv{\sigma}} = f\left((\sigma^{(1)}_1, \dots, \sigma^{(1)}_{\sscR})_2, \dots, (\sigma^{(\sscN)}_1, \dots, \sigma^{(\sscN)}_{\sscR})_2\right),
\end{align}
where $\vv{\sigma}$ denote the set of all binary indices.
The total number of tensor elements is $2^\scL$, which is computationally intractable even for a moderate value of  $\cL$.

To address this, we approximate the tensor $F_{\vv{\sigma}}$ in the Tensor Train (TT) format:
\begin{align}
    F_{\vv{\sigma}} \simeq  \widetilde{F}_{\vv{\sigma}} = \sum_{{\alpha_1}}\cdots\sum_{\alpha_{\sscL-1}}
F^{(1)}_{\alpha_0 \sigma_1 \alpha_1}
F^{(2)}_{\alpha_1 \sigma_2 \alpha_2}
\cdots
F^{(d)}_{\alpha_{\sscL-1} \sigma_{\sscL} \alpha_{\sscL}},
\end{align}
where each $F^{(\ell)}$ is a three-way tensor (TT-core) of size $\chi_{\ell-1}\times 2 \times \chi_\ell$, and $\chi_\ell$ is referred to as the bond dimension. The memory cost scales as $\mathcal{O}(\chi^2\cL)$, where $\chi = \max_{\ell}\chi_\ell$. If $f(\vv{u})$ has a low-rank structure, then $\chi$ remains small and independent of $\cL$, allowing efficient representation of exponentially large tensors.
To construct this TT efficiently, we use TCI~\cite{fernandez2024learning}, an adaptive algorithm that selects only a small subset of the full tensor entries. The bond dimensions $\chi_\ell$ are automatically optimized to satisfy $\| F - \widetilde{F} \|_{\mathrm{max}} / \|F\|_{\mathrm{max}} < \tau$, where $\tau$ is a prescribed tolerance, $\widetilde{F}$ denotes the TT approximation, and $\|\cdot\|_{\mathrm{max}}$ is the maximum norm.
For the sampling strategy, we adopt the partial rank-revealing LU decomposition, combined with the local-update and global-update scheme~\cite{ishida2024low} to ensure numerical stability and accuracy. The number of required function evaluations typically scales as $\mathcal{O}(\chi^2\cL)$, which remains practical as long as $f(\vv{u})$ exhibits a low-rank structure.

\subsection{Integration in quantics tensor train format}

Once the discretized tensor is constructed in the tensor train format $\widetilde{F}_{\vv{\sigma}}$, the integration of $f(\vv{u})$ over the domain $[0, 1)^\sscN$ can be efficiently approximated by a summation over all binary indices:
\begin{align}
    \int_{[0, 1)^\sscN} d\vv{u}\, f(\vv{u}) \simeq \dfrac{1}{2^{\sscL}}\sum_{\vv{\sigma}}F_{\vv{\sigma}}.
\end{align}
This approximation corresponds to a Riemann sum over a uniform grid with $2^\sscL$ points. Although Riemann summation is typically considered inefficient for high-dimensional integrals, it becomes highly effective in this framework due to a key property: each grid point is defined via a binary expansion, and the resulting exponentially large tensor can be efficiently compressed when the integrand has a low-rank structure. 

The summation over all binary indices can be performed directly using the TT representation:
\begin{align}
    \sum_{\vv{\sigma}}F_{\vv{\sigma}} 
    &= \sum_{\vv{\alpha}}\left(\sum_{\sigma_1=0}^1 F^{(1)}_{\alpha_0\sigma_1\alpha_1}\right)\cdots\left(\sum_{\sigma_{\sscL}=0}^1 F^{(\scL)}_{\alpha_{\sscL-1}\sigma_{\sscL}\alpha_{\sscL}}\right),
\end{align}
where $\vv{\sigma}$ and $\vv{\alpha}$ are the index vectors $(\sigma_1, \cdots, \sigma_{\sscL})$ and $(\alpha_1, \cdots, \alpha_{\sscL-1})$, with boundary conditions $\alpha_0 = \alpha_\sscL = 1$.
Each inner sum over $\sigma_\ell\in\{0, 1\}$ corresponds to contracting the TT core $F^{(\ell)}$ with the vector $(1,1)$, which can be interpreted as summing over the binary basis. The full contraction proceeds sequentially, with total computational cost scaling as $\mathcal{O}(\chi^3\cL)$. This structure enables exponentially fine discretization while maintaining linear memory and computational costs with respect to $\cL$. As a result, the Riemann sum, although often considered naive or inefficient, becomes a highly competitive integration method for functions with low-rank structure. For more details on this method, we refer the reader to see the supplemental materials of Ref.~\onlinecite{ishida2024low}.

\section{result}\label{sec-result}
This section shows our numerical results of the AC and DC spin currents.
In a numerical result, we will see that the $x$ and $y$-spin-polarized AC spin current has a non-zero value. In contrast, the $z$-spin-polarized DC spin current has a non-zero value, where the spin of FMI has precessional motion around the $z$-axis.
First, we present the results of the AC spin current, followed by the results of the DC spin current.
In each section, we show the temperature, frequency, and thickness dependence of the spin current.

\begin{figure}[ht]
    \centering
    \subfigure[$J_{\mathrm{s},x}^{\rm AC, norm}(L,\Omega)$]{
    \includegraphics[scale=0.7]{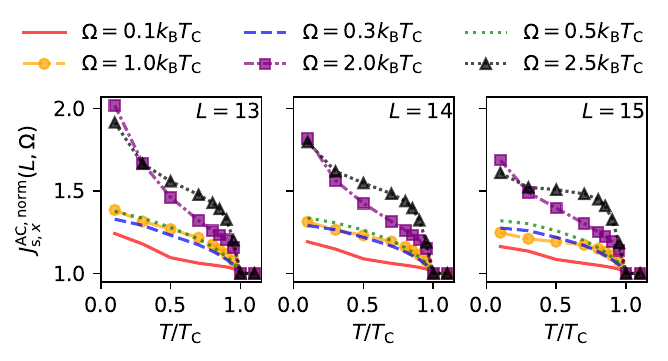}
    \label{fig:temp-Jx-AC}
    }
    \subfigure[$J_{\mathrm{s},y}^{\rm AC, norm}(L,\Omega)$]{
    \includegraphics[scale=0.7]{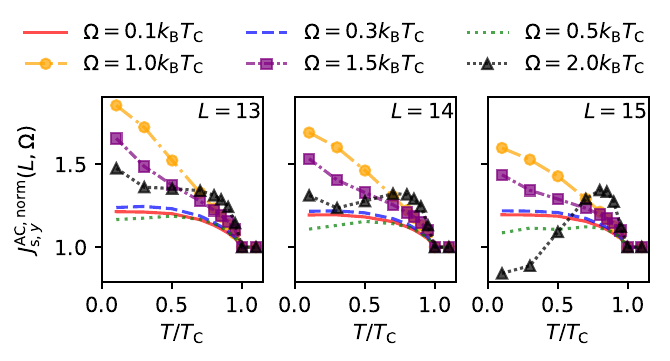}
    \label{fig:temp-Jy-AC}
    }
    \caption{The normalized AC spin current defined on Eq.~(\ref{J-AC-norm}).
    The parameters are set as $t=4.0 \Delta_{0}(0)$, $T_{\rm C} = 1.0$ K and $\mu = -0.2 t$.
    Each panels correspond to $L=12$, $13$, and $14$.
    (a) shows the $x$-component of the AC spin current.
    The solid (red), dashed (blue), and dotted (green) lines without markers correspond to $\Omega/ k_{\rm B}T_{\rm C} = 0.1$, $0.3$, and $0.5$, respectively.
    The dashed (circle), dotted (square) and dotted (triangle) with markers correspond to $\Omega/ k_{\rm B}T_{\rm C} = 1.0$, $1.5$, and $2.5$, respectively.
    (b) shows the $y$-component of the AC spin current.
    The solid (red), dashed (blue), and dotted (green) lines without markers correspond to $\Omega/ k_{\rm B}T_{\rm C} = 0.1$, $0.3$, and $0.5$, respectively.
    The dashed (circle), dotted (square) and dotted (triangle) with markers correspond to $\Omega/ k_{\rm B}T_{\rm C} = 1.0$, $1.5$, and $2.0$, respectively.
    The $z$-component of the AC spin current is always zero.
    }
    \label{fig:temp-J-AC}
\end{figure}

\subsection{AC spin current}
Here, we show the numerical results of the AC spin current defined in Eq.~(\ref{AC-spin-current}).
\subsubsection{Temperature dependence}

FIG.~\ref{fig:temp-J-AC} shows the temperature dependence of the AC spin current.
The $x$-component of the AC spin current is shown in FIG.~\ref{fig:temp-Jx-AC}, and the $y$-component is shown in FIG.~\ref{fig:temp-Jy-AC}.
From these figures, we can observe similar behaviors of the $x$ and $y$-components of the AC spin current.
Both components of the AC spin current enhanced quickly at the critical temperature $T_{\rm C}$, and increase monotonically along with lowering the temperature.
This behavior is attributed to the coherence of the superconducting state, specifically corresponding to the coherence peak.
However, there is a slight difference between the $x$ and $y$-components at high microwave frequencies.
In this region, the $y$-component is suppressed in the low temperature region with a thin layer.
This difference may be caused by the finite size effect.

\subsubsection{Frequency dependence}
\begin{figure}
    \centering
    \subfigure[$x$-component of AC spin current]{
        \includegraphics[scale=0.7]{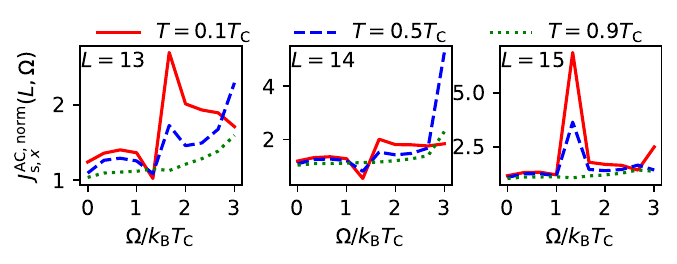}
        \label{fig:AC-freq-x}
    }

    \subfigure[$J_{\mathrm{s},y}^{\rm AC, norm}(L,\Omega)$]{
        \includegraphics[scale=0.7]{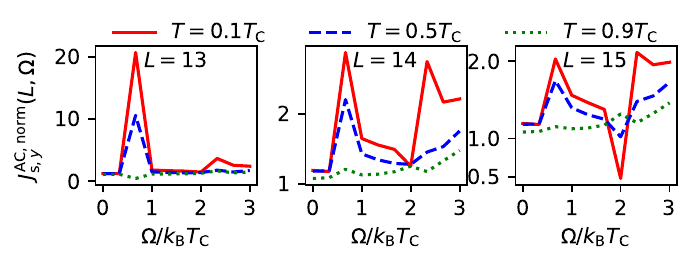}
        \label{fig:AC-freq-y}
    }
    \caption{Frequency dependence of the AC spin current.
    The parameters are set as $t=4.0 \Delta_{0}(0)$, $T_{\rm C} = 1.0$ K and $\mu = -0.2 t$.
    The solid (red), dashed (blue), and dotted (green) lines correspond to $L = 12$, $13$, and $14$, respectively.
    (a) shows the $x$-component of the AC spin current.
    (b) shows the $y$-component of the AC spin current.
    }
\end{figure}

Next, we show the frequency dependence of the AC spin current.
FIG.~\ref{fig:AC-freq-x} and \ref{fig:AC-freq-y} show the $x$- and $y$-components, respectively.
First, we can see the singular behavior of $J_{\mathrm{s},x}^{\rm AC, norm}(L,\Omega)$ at $1.0 < \Omega/k_{\rm B} T_{\rm C}< 2.0$.
However, the $y$-component have singularities in $0.0 < \Omega/k_{\rm B} T_{\rm C}< 1.0$ and $\Omega/k_{\rm B}T_{\rm C} \simeq 2.0$.
These singularities may be understood as the ``resonance'' between the microwave frequency and the superconducting gap.
However, the reasons why the singularities of the $x$ and $y$-components are different are unclear.
This point will be discussed at the end of this section.

\subsubsection{Thickness dependence}
\begin{figure}
    \centering 
    \subfigure[$x$-component of AC spin current]{
        \includegraphics[scale=0.7]{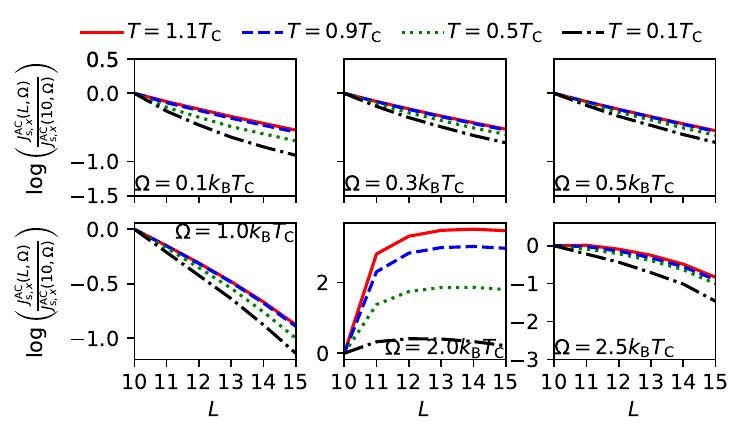}
        \label{fig:AC-thickness-x}
    }
    \subfigure[$y$-component of AC spin current]{
        \includegraphics[scale=0.7]{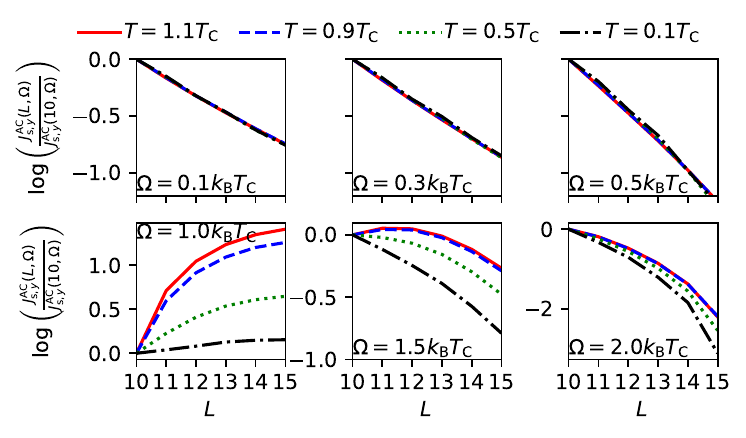}
        \label{fig:AC-thickness-y}
    }
    \caption{Thickness dependence of the AC spin current.
    The parameters are set as $t=4.0 \Delta_{0}(0)$, $T_{\rm C} = 1.0$ K and $\mu = -0.2 t$.
    The $L$ is thickness of the superconducting layer, and the legend shows the microwave frequencies normalized by the critical temperature $T_{\rm C}$.
    The solid (red), dashed (blue), dotted (green) and dashdot (black) lines correspond to $T/T_{\rm C} = 1.1$, $0.9$, $0.5$ and $0.1$, respectively.
    (a) shows the $x$-component of the AC spin current.
    The panels shown the frequency $\Omega/k_{\rm B}T_{\rm C} = 0.1$, $0.3$, $0.5$, $1.0$, $2.0$, and $2.5$.
    (b) shows the $y$-component of the AC spin current.
    The panels shown the frequency $\Omega/k_{\rm B}T_{\rm C} = 0.1$, $0.3$, $0.5$, $1.0$, $1.5$, and $2.0$.
    }
    \label{fig:AC-thickness}
\end{figure}

This section shows the thickness dependence of the AC spin current.
To analyze the thickness dependence, we consider the normalization of the AC spin current defined as 
\begin{align}
     \frac{J_{\mathrm{s},\alpha}^{\rm AC}(L,\Omega)}{J_{\mathrm{s},\alpha}^{\rm AC}(L=10,\Omega)}.
    \label{J-AC-norm-L}
\end{align}
In the FIG.~\ref{fig:AC-thickness}, we show the $x$ and $y$-components of the AC spin current.
The qualitative behavior of the $x$ and $y$-components is similar.
The AC spin current decays exponentially with increasing thickness $L$ at low microwave frequencies.
Furthermore, there exists a singular frequency increasing with thickness, and the decay is no longer exponential after that singular frequency. 
However, the value of the singular frequency is different between the $x$ and $y$-components.
Additionally, quantitatively, the $x$-component tends to have a more pronounced effect on the superconducting state in the low frequency range compared to the $y$-component.
The reasons why the $x$ and $y$-components have different behaviors and exhibit the singular frequency may be understood as follows.

\subsubsection{Discussion on the AC spin current}

Our numerical results reveal a crossover behavior in the thickness dependence of the AC spin current: at low microwave frequencies, the current decays exponentially with increasing SC layer thickness, while at higher frequencies, this behavior transitions to a non-monotonic, oscillatory profile.

We attribute this crossover to interference effects of quasiparticles within the finite-length SC layer.
When the SC thickness becomes comparable to the quasiparticle coherence length, multiple reflections at the FMI/SC and SC/NM interfaces can lead to interference phenomena analogous to Fabry–Pérot-type resonances \cite{Fabry-Perot-diodes-2024,Fabry-Perot-resonances-2010}.
Indeed, such interference effects have been reported in theoretical studies of spin and charge transport in superconductor/ferromagnet hybrid systems, where oscillatory behavior in the current as a function of SC layer thickness was found to result from quasiparticle phase coherence and Andreev-type reflections \cite{triplet-induced-swave-2008,linder2015superconducting,sonin2010spin,spin-current-interference-2008}.
This mechanism can modulate the transmission amplitude of spin-polarized carriers through the SC depending sensitively on both the excitation frequency and the layer thickness.
Notably, these resonances can occur even above the superconducting transition temperature, indicating that their origin lies in quasiparticle coherence rather than superconductivity itself.

Moreover, the observed asymmetry between the $x$- and $y$-components of the AC spin current, especially in the high-frequency regime, indicates a spin-dependent interference mechanism. 
This could stem from spin-mixing effects at the interfaces, where the spin polarization vector experiences component-dependent phase shifts. 
These phase shifts alter the conditions for constructive interference differently for each spin component.

While our current model captures these features numerically, a full analytical treatment that explains the component-selective resonance remains a challenging open problem. Further insight may be gained through analytical studies of simplified toy models or by evaluating spatially resolved spin accumulation profiles.

\subsection{DC spin current}
In this section, we show the results of the DC spin current defined in Eq.~(\ref{pumped-spin-current-simp}).

\subsubsection{Temperature dependence}
\begin{figure}
    \centering
    \includegraphics[scale=0.7]{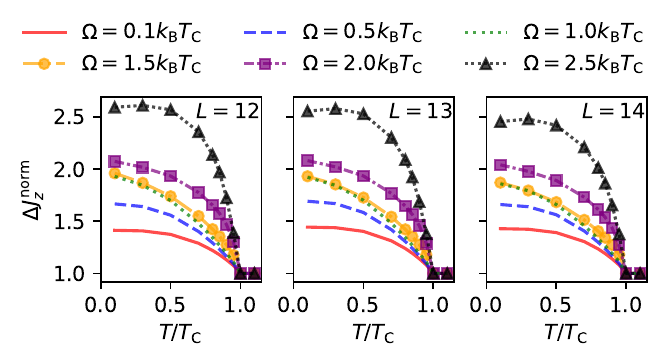}
    \caption{The normalized DC spin current defined on Eq.~(\ref{J_renom-T}).
    The parameters are set as  $t=4.0 \Delta_{0}(0)$, $T_{\rm C} = 1.0$ K and $\mu = -0.2 t$.
    The solid (red), dashed (blue), and dotted (green) lines without markers correspond to $\Omega/ k_{\rm B}T_{\rm C} = 0.1$, $0.5$, and $1.0$, respectively.
    The lines with circles (yellow), squares (purple), and triangles (black) correspond to $\Omega/ k_{\rm B}T_{\rm C} = 1.5$, $2.0$, and $2.5$, respectively.
    The $L$ is thickness of the superconducting layer, and the legend shows the microwave frequencies normalized by the critical temperature $T_{\rm C}$.
    In numerical, we get $\Delta J_{x} = \Delta {J}_{y} = 0$.}
    \label{fig:T-spin}
\end{figure}

FIG.~\ref{fig:T-spin} shows the temperature dependence of the pumped DC spin current.
This figure shows the $z$-component of normalized DC spin current defined by (\ref{J_renom-T}).
The most typical feature of the DC spin current is to have a finite value of the $z$-component, and the $x$ and $y$-components are always zero.
The reasons why this difference occurs are that the magnetization of the FMI makes a steady precessional motion around the $z$-axis due to the circularly polarized microwave.
Therefore, the $z$-component and the $x$- ($y$-) component of the pumped spin current become DC and AC spin current, respectively.

According to the FIG,~\ref{fig:T-spin}, the DC spin current is monotonically increasing with decreasing temperature at $T < T_{\rm C}$ for all thickness and frequencies condition.
Therefore, it can be concluded that the DC spin current pumped in the NM layer is enhanced by the superconducting state.

\subsubsection{Frequency dependence}
\begin{figure}[b]
    \centering
    \includegraphics[scale=0.7]{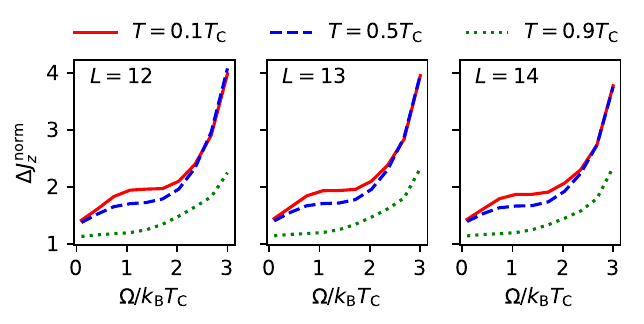}
    \caption{Frequency dependence of the DC spin current.
    The parameters are set as $t=4.0 \Delta_{0}(0)$, $T_{\rm C} = 1.0$ K and $\mu = -0.4 t$.
    The solid (red), dashed (blue), and dotted (green) lines correspond to $T/T_{\rm C} = 0.1$, $0.5$, and $0.9$, respectively.
    Each panels is data for $L = 12$, $13$, and $14$.
    }
    \label{fig:freq-DC-spin}
\end{figure}

This section shows the frequency dependence of the DC spin current (FIG.~\ref{fig:freq-DC-spin}).
We can see that the DC spin current is monotonically increasing with increasing frequency for all thickness and temperature conditions.
However, there is a plateau in the middle frequency region.
This plateau could be generated by the following scenario.
The DC spin current is transported by both the spin fluctuation of the ground state and quasiparticle excitations.
The spin fluctuation of the ground state can occur by an infinitesimal microwave frequency.
On the other hand, the quasiparticle excitations need a microwave frequency that is large enough to overcome the superconducting gap.
Therefore, the DC spin current increases in the low frequency region and again in the high frequency region after the plateau.

\subsubsection{Thickness dependence}
\begin{figure}
    \raggedright
    \includegraphics[scale=0.69]{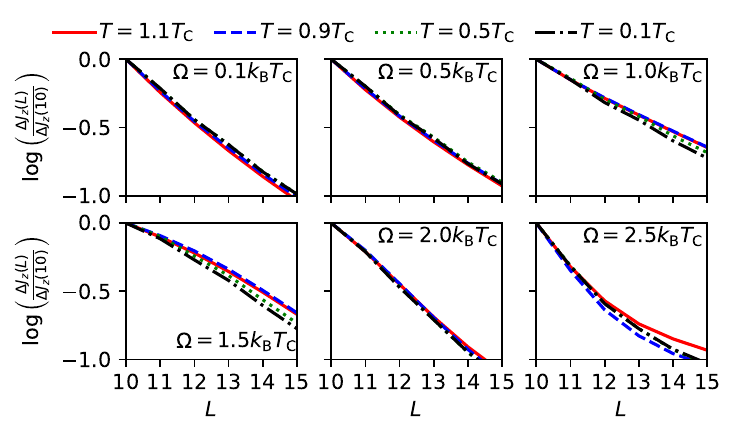}
    \caption{The Thickness of the superconductor $L$ dependence of the DC spin current.
    In this figure, we always set $t=4.0 \Delta_{0}(0)$, $T_{\rm C} = 1.0$ K and $\mu = -0.4 t$.
    Panels show the thickness dependence with $L$ and $\Omega/k_{\rm B} T_{\rm C} = 0.1$, $0.5$, $1.0$, $1.5$, $2.0$, and $2.5$. 
    The solid (red), dashed (blue), dotted (green), and dashdot (black) lines correspond to $T/T_{\rm C} = 1.1$, $0.9$, $0.5$ and $0.1$, respectively.
    }
    \label{fig:L-DC-spin}
\end{figure}

We now examine the dependence of the spin current on the thickness $L$ of the superconducting layer and on the microwave frequency $\Omega$.
To analyze the thickness dependence, we apply an alternative normalization of the spin current as defined in Eq.~(\ref{pumped-spin-current-simp}).
Specifically, we consider the ratio
\begin{align}
    \frac{\Delta J^{(3)}_{\mathrm{s}, i}(L,\omega_{\rm s})}{\Delta J^{(3)}_{\mathrm{s}, i}(L_{0},\omega_{\rm s})}
    = \frac{\mathcal{J}_{\mathrm{s}, i}(L)}{\mathcal{J}_{\mathrm{s}, i}(L_{0})},
\end{align}
where $L_{0}$ is a reference value.
FIG.~\ref{fig:L-DC-spin} shows the thickness dependence of the DC spin current for $\Omega/k_{\rm B} T_{\rm C} = 0.1$, $0.5$, $1.0$, $1.5$, $2.0$, and $2.5$.
We can observe that the DC spin current decays exponentially with increasing thickness $L$ in the low-frequency region.
On the other hand, in the high-frequency region, the dependency is qualitatively different.
Thus, this result supports the scenario that the DC spin current is transported by both the spin fluctuation of the ground state and quasiparticle excitations.

Furthermore, in the low-frequency region, the decay rate of the DC spin current at low temperatures is almost the same before the superconducting transition. 
While in the high frequency region, the decay rate of the DC spin current at low temperatures increases with the superconducting transition.
This indicates that the quasiparticle is disadvantageous for the DC spin current transportation in the superconducting state.

\subsection{Technical attention}
In the present case, the dependence of the integrand function for our results on the TT can be explicitly written as $f(\nu, k_x, k_y, \epsilon)$, where $\nu=x, y, z$ denotes the component of the spin current vector. To make the integration range of $\epsilon$ finite, we perform a variable transformation $\epsilon\to\tan(\pi\epsilon/2)$. Hereafter, we denote the transformed variable as $\epsilon$.
The continuous variables $(k_x, k_y, \epsilon)$ are discretized using the quantics representation. The number of bits for the $k$-vector is set to $\cR_k = 6$, and for the energy to $\cR_\epsilon = 15$, chosen to ensure sufficient convergence of the integration. 

We now construct the TT of the integrand by using TCI as typical parameter sets. The tolerance of TCI is set to $\tau=10^{-6}$. The TCI construction takes ten minutes with a single core and thread for $\chi \simeq 300$. The subsequent summation over the internal variables can be performed instantaneously.

\section{Conclusion}\label{sec-conclusion}
In this paper, we investigate spin pumping in a trilayer system consisting of FMI/SC/NM. 
We have derived the AC and DC spin current injected into the NM layer due to the ferromagnetic resonance in the FMI layer under circularly polarized microwave irradiation.
While the AC spin current was derived by treating spin motion in FMI as a classical given value, the DC spin current was obtained by treating the spin motion as a quantum quasiparticle (i.e. uniform magnons) within the Keldysh Green’s function framework.

We first presented the temperature dependence of the spin current.
Both AC and DC spin currents exhibit a coherence peak below the superconducting critical temperature.
While the DC spin current increases monotonically with lowering temperature, we observe that the specific case of the AC spin current can decrease with decreasing temperature.
We also investigated the frequency dependence of the spin current.
The DC spin current shows a monotonic increase with increasing microwave frequency, while the AC spin current exhibits a singular behavior.
Furthermore, the AC and DC spin currents have exponentially decaying behavior with increasing thickness of the SC layer at low microwave frequencies.
What both AC and DC spin current have in common is that there is a certain characteristic frequency above which they cease to decay exponentially.
However, this tendency appeared stronger for the AC case.
The above DC spin current behavior can be understood by considering that the ground state of the superconducting state contributes at low frequencies, while the spin current is transported by quasiparticle excitation in the high-frequency region.
However, the origin of the singular behavior in the AC spin current, especially the polarization-dependent resonance features, is not fully understood within the current framework. 
We consider this an interesting direction for future investigation.

The theoretical framework developed here provides a powerful tool for analyzing spin pumping phenomena in various multilayer systems.
For example, spin pumping through unconventional superconductors such as $p$-wave, $d$-wave, and topological superconductors can be investigated using the same formalism.
Moreover, the variation for the third layer is also possible and interesting.
For instance, the effects of the spin-orbit coupling in the NM layer or at its surface can be incorporated into the model.
From a more modern perspective, spin pumping using antiferromagnet or altermagnet is another important and interesting topic.

\begin{acknowledgments}
 This work is supported by JST SPRING (Grant No. JPMJSP2125). H. I. is supported by JST FOREST (Grant No. JPMJFR2232).
\end{acknowledgments}

\appendix

\section{Derivation of the AC spin current}\label{app-AC}
In this section, we derive the AC spin current shown in Eq.~(\ref{AC-spin-current}).
The starting point is the definition of the spin current in the NM layer, which is given in Eq.~(\ref{def-spin-current}).
Now, we consider the theory for the expectation value in the definition of the spin current as the action in Eq.~(\ref{S-AC}).
As the perturbation expansion of the interaction term $S_{\rm int}^{\rm cl}$, we consider the first order perturbation of $V$ and $T_{\rm int}$, and we can evaluate the expectation value as
\begin{align}
    &\expval{ c_{L\sigma',\vv{k}}^{\dag}(t)  \vec{\tilde{\sigma}}^{\sigma'\sigma} d_{L-1\sigma,\vv{k}}(t)} 
    \notag\\ 
    &\simeq \int_{\mathcal C} \frac{dt_1 dt_2}{2}
    \sum_{\alpha=\pm}  \sum_{\sigma_1,\sigma_2,\sigma_2'} 
    S_{\alpha}(t_2)
    \notag\\
    &\times \expval{\tau_{z} c_{L\sigma_{1}\vv{k}}(t_1) c_{L\sigma',\vv{k}}^{\dag}(t)} 
    \vec{\tilde{\sigma}}^{\sigma'\sigma}
    \bar{\sigma}_{-\alpha}^{\sigma_2 \sigma_2'}
    \notag\\ 
    &\times 
    \expval{ d_{L-1 \sigma,\vv{k}}(t) d_{0\sigma_2',\vv{k}}(t_2) \bar{d}_{L-1 \sigma_1,\vv{k}}(t_1) \bar{d}_{0\sigma_2,\vv{k}}(t_2) }
    \notag\\ 
    &= \int_{\mathcal C} \frac{dt_1 dt_2}{2}
    \sum_{\alpha=\pm}  \sum_{\sigma_1,\sigma_2,\sigma_2'}
    S_{\alpha}(t_2)
    \vec{\tilde{\sigma}}^{\sigma'\sigma}
    \notag\\
    &\times \expval{ d_{L-1 \sigma,\vv{k}}(t) \bar{d}_{0\sigma_2,\vv{k}}(t_2) }
    \bar{\sigma}_{-\alpha}^{\sigma_2 \sigma_2'}
    \expval{ d_{0\sigma_2',\vv{k}}(t_2) \bar{d}_{L-1 \sigma_1,\vv{k}}(t_1) }
    \notag\\
    &\times \tau_{z}\expval{ c_{L\sigma_{1}\vv{k}}(t_1) c_{L\sigma',\vv{k}}^{\dag}(t)}
    \notag\\ 
    &= -\frac{i}{2} \int_{\mathcal C} dt_1 dt_2
    \sum_{\alpha=\pm}  \sum_{\sigma_1,\sigma_2,\sigma_2'}
    S_{\alpha}(t_2)
    \vec{\tilde{\sigma}}^{\sigma'\sigma}
    \notag\\
    &\times G_{\rm SC}^{L-1 \sigma, 0\sigma_2}(\vv{k},t,t_2)
    \bar{\sigma}_{-\alpha}^{\sigma_2 \sigma_2'}
    G_{\rm SC}^{0\sigma_2', L-1 \sigma_1}(\vv{k},t_2,t_1)
    \notag\\
    &\times
    \tau_{z} G_{\rm NM}^{L\sigma_1, L\sigma'}(\vv{k},t_1,t)
\end{align}
where $\int_{\mathcal C} dt$ denotes the contour-ordered integration with $\mathcal{C}: -\infty \to +\infty \to -\infty$, and we use the Wick's theorem to split the expectation value with four fermion fields into it with two fermion fields.
Therefore, the spin current is expressed as
\begin{align}
    \vec{J}_{\rm s}(L,t)
    &= -\frac{T_{\rm int}^2 V}{8}\sum_{\alpha=\pm}  \sum_{\sigma_1,\sigma_2,\sigma_2'}\sum_{\sigma,\sigma'}\sum_{\vv{k} \in \rm BZ}
    \int_{\mathcal C} dt_1 dt_2
    \notag\\ 
    &\times
    \Re
    \left[
    S_{\alpha}(t_2) 
    \vec{\tilde{\sigma}}^{\sigma'\sigma}
    G_{\rm SC}^{L-1 \sigma, 0\sigma_2}(\vv{k},t,t_2)
    \bar{\sigma}_{-\alpha}^{\sigma_2 \sigma_2'}
    \right.
    \notag\\
    &\left.\times
    G_{\rm SC}^{0\sigma_2', L-1 \sigma_1}(\vv{k},t_2,t_1)
    \tau_{z} G_{\rm NM}^{L\sigma_1, L\sigma'}(\vv{k},t_1,t)
    \right].
\end{align}
Here, we consider the case of circularly polarized light, where the spin field is given by $S_{\alpha}(t) = S e^{-i\alpha \Omega t}$.
Then, the Fourier component of the spin current $\vec{J}_{\rm s}(L,\omega) := \int dt e^{i\omega t}\vec{J}_{\rm s}(L,t)$ is given by Eq.~(\ref{AC-spin-current}) with the Langreth rules.

\section{Green's function of the resonanced magnon}\label{app-mag}
Here, we derive the correction terms to the magnon Green's function due to the resonance with the circular polarized magnetic field.
We base on the system defined by the Hamiltonian in Eq.~(\ref{H-Mag}).
Now, we consider the Green's function of the magnon defined as 
\begin{align}
    G_{\rm M}(\vv{q},t_1,t_2) =-i \expval{\mathcal{T}_{\mathcal C} a_{\vv{q}}(t_1) a_{\vv{q}}^{\dag}(t_2)}_{\rm M},
\end{align}
where the expectation value is calculated along with the theory defined by the Hamiltonians in Eq.~(\ref{H-Mag}).
The correction to the Green's function by the second-order perturbation expansion is written as
\begin{align}
    &\Delta G(\vv{q},t_1,t_2)
    \notag\\
    &= -\frac{i}{2}\expval{ \int_{\mathcal C}dt dt' a_{\vv{q}}(t_1) a_{\vv{q}}^{\dag}(t_2)H_{\rm M-h}(t)H_{\rm M-h}(t') }_{0,\rm M}
    \notag\\ 
    &= -i\lambda^2 \expval{ \int_{\mathcal C}dt dt' a_{\vv{q}}(t_1) a_{\vv{q}}^{\dag}(t_2)h_{+}(t) a_{\vv{0}}^{\dag}(t) h_{-}(t') a_{\vv{0}}(t') }_{0,\rm M},
\end{align}
where the integral $\int_{\mathcal C}dt$ is the contour-orderd integration.
According to Wick's theorem, we can get only one disconnected term as
\begin{align}
    \Delta G(\vv{q},t_1,t_2)
    &= -i \delta_{\vv{q},\vv{0}} \lambda^2 \lb G_{\rm M}^{(0)} \star \Sigma_{\rm h} \star G_{\rm M}^{(0)} \rb (\vv{q},t_1,t_2),
    \\
    \Sigma_{\rm h}(t,t') &= \expval{\mathcal{T}_{\mathcal C} h_{+}(t) h_{-}(t')}.
\end{align}
From the definitions of the retarded and advanced Green's function, we can find $\Sigma_{\rm h}^{\rm R}(t,t') = \Sigma_{\rm h}^{\rm A}(t,t') = 0$.
Therefore, the correction term is written as 
\begin{align}
    &\Delta G^{\rm R}(\vv{0},t_1,t_2) = \Delta G^{\rm A}(\vv{0},t_1,t_2) = 0,
    \notag\\
    &\Delta G^{<}(\vv{0},t_1,t_2)
    \notag\\
    &= -i \lambda^2 \int dtdt' G_{\rm M}^{(0)\rm R}(\vv{0},t_1,t)  \Sigma_{\rm h}^{<}(t,t') G_{\rm M}^{(0)\rm A}(\vv{0},t',t_2)
\end{align}
If we define the Fourier transformation along with the time as $G_{\rm M}^{(0)\rm R, A,<}(\vv{0},t_1,t) = (2\pi)^{-1/2}\int d\omega G_{\rm M}^{(0)\rm R, A,<}(\vv{0},\omega) e^{i\omega (t_2 - t_1)} $, then the correction term is found as
\begin{align}
    \Delta G^{<}_{\rm M}(\vv{q},\omega) = -2i\pi \lambda^2 h^2 \delta_{\vv{q},\vv{0}} \delta(\omega - \Omega) \abs{G_{\rm M}^{(0)\rm R}(\vv{q},\omega)}^2
    \label{app-Mag-correc}
\end{align}

\section{Derivation of the DC spin current}\label{app-spin-current}
\begin{figure}
    \centering
    \includegraphics[scale=0.8]{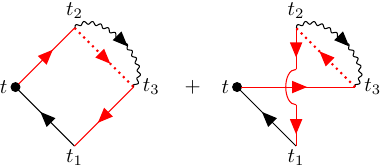}
    \caption{Feynman diagrams of connected terms of spin current $\vec{J}_{\rm s, con}^{(3)}(L,t,t_1,t_2,t_3)$.
    The black solid lines and wavy lines represent the normal metal Green's function $G_{\rm NM}(\vv{k},t_1,t_2)$ and the magnon Green's function $G_{\rm M}(\vv{q},t_2,t_3)$, respectively.
    The red solid and dashed lines represent the superconducting Green's function $G_{\rm SC}(\vv{k},t_1,t_2)$ of which endpoints connected to the normal metal and magnon Green's function correspond to the $L-1$th and $0$th layer degrees of freedom, respectively.
    The black vertex at time $t$ denote the product of $\vec{\tilde{\sigma}}$.
    }
    \label{fig:time-dia}
\end{figure}

This section derives the DC spin current by performing the perturbation expansion of the spin current defined in Eq.~(\ref{def-spin-current}) and its operator representation is seen in Eq.~(\ref{J-spin-ope}).
Since we want to derive the spin current as a reaction to the microwave, we must consider the second-order perturbation to include an even number of magnon fields.
However, the second-order perturbation theory includes only one electron field $c_{L\sigma',\vv{k}}^{\dag}(t)$.
Therefore, the lowest-order perturbation in our purpose is the third-order terms of the perturbation expansion written as
\begin{align}
    &\vec{J}_{\rm s}^{(3)}(L,t)
    = -\frac{T_{\rm int}^2 V^{2}}{6}\sum_{\sigma,\sigma'}\sum_{1,2,3}
    \lim_{t'\to t}
    \int_{\mathcal{C}} dt_1 dt_2 dt_3 
    \notag\\ 
    &\times
    \Re \Tr \LB  -\vec{\chi}_{\rm SC}(\vv{k},\vv{q},t,t',t_2,t_3) G_{\rm M}(\vv{q},t_2,t_3) \RB,
    \\ 
    &\vec{\chi}_{\rm SC}(\vv{k},\vv{q},t,t',t_2,t_3)
    \notag\\
    &:= G_{\rm NM}(\vv{k},t_1,t_2) \vec{\tilde{\sigma}}
    \lb \bar{\sigma}_{-}^{\sigma_2'\sigma_2} \rb
    \lb \bar{\sigma}_{+}^{\sigma_3\sigma_3'} \rb
    \notag\\ 
    &\times \langle
    T_{\mathcal{C}} 
    d_{L-1\sigma,\vv{k}}(t') d_{0\sigma_{2},\vv{k}_2}(t_2) d_{0\sigma_{3}',\vv{k}_3+\vv{q}}(t_3)
    \notag\\
    &\times
    d_{L-1\sigma_{1},\vv{k}}^{\dag}(t_1) d_{L-1\sigma_{2}',\vv{k}_2+\vv{q}}^{\dag}(t_2) d_{0\sigma_{3},\vv{k}_3}(t_3)
    \rangle_{0},
    \\ 
    &G_{\rm NM}(\vv{k},t_1,t_2) := -i \expval{T_{\mathcal{C}} c_{L\sigma_{1},\vv{k}}(t_1) c_{L\sigma',\vv{k}}^{\dag}(t) }_{0},
\end{align}
where $\sum_{1,2,3}$ denotes the summation over the permutations of the indices $\sigma_i$ and $\sigma_{i}'$ ($i = 1,2,3$). 
According to Wick's theorem in $\vec{\chi}_{\rm SC}(\vv{k},\vv{q},t,t',t_2,t_3)$, the connected terms are written as 
\begin{align}
    &\vec{J}_{\rm s}^{(3)}(L,t)
    \notag\\
    &=
    \frac{T_{\rm int}^2 V^{2}}{6} \lim_{t'\to t}
    \Re \Tr
    \notag\\  
    &\times\left[
        \vec{\chi}_{1}(t,t_2) \star G_{\rm M}(\vv{0},t_2,t_3) \star \lb \bar{\sigma}_{+} G_{\rm SC}^{0\sigma_3',0\sigma_3}(t_3,t_3) \rb
    \right.
    \notag\\
    &+
    \lb  G_{\rm SC}^{0\sigma_2,0\sigma_2'}(t_2,t_2) \bar{\sigma}_{-} \rb \star G_{\rm M}(\vv{0},t_2,t_3) \star \vec{\chi}_{2}(t_3,t) 
    \notag\\
    &-\left. \int_{\mathcal{C}}dt_1 \vec{J}_{\rm s, con}^{(3)}(L,t,t_1,t_2,t_3) \right],
\end{align}
\begin{align}
    &\vec{J}_{\rm s}^{(3)}(L,t)
    \notag\\
    &=
    \frac{T_{\rm int}^2 V^{2}}{6} \lim_{t'\to t}
    \Re \Tr
    \notag\\  
    &\times\left[
        \lB\vec{\chi}_{1}\star G_{\rm M}(\vv{0}) \star \lb \bar{\sigma}_{+} G_{\rm SC}^{0\sigma_3',0\sigma_3}(\vv{0})\rb\rB(t,t')
    \right.
    \notag\\
    &+
    \lB\lb  G_{\rm SC}^{0\sigma_2,0\sigma_2'} \bar{\sigma}_{-} \rb \star G_{\rm M}(\vv{0}) \star \vec{\chi}_{2} \rB(t',t)
    \notag\\
    &-\left. \int_{\mathcal{C}}dt_1dt_2 dt_3 \vec{J}_{\rm s, con}^{(3)}(L,t,t_1,t_2,t_3) \right],
\end{align}
where the $\vec{J}_{\rm s, con}^{(3)}(L,t,t_1,t_2,t_3)$ denote the contribution of connected diagrams shown in FIG.~\ref{fig:time-dia}, and the $\star$ denotes the convolution (Moyal product) defined as
\begin{align}
    \lB G \star G' \rB(t,t') := \int_{\mathcal{C}} dt_1 G(t,t_1) G'(t_1,t').
\end{align}
In the disconnected terms, the $\vec{\chi}_{1}(t,t_2)$ and $\vec{\chi}_{2}(t_3,t)$ are defined as
\begin{align}
    &\vec{\chi}_{1}(t,t_2)
    \notag\\
    &= -i\Tr \int_{\mathcal{C}} dt_1
    G_{\rm NM}^{L\sigma_1,L\sigma'}(\vv{k},t_1,t)
    \vec{\tilde{\sigma}}^{\sigma'\sigma}
    G_{\rm SC}^{L-1\sigma,0\sigma_{2}'}(\vv{k},t',t_2)
    \notag\\
    &\times
    \tau_e\sigma_{-}^{\sigma_{2}',\sigma_{2}}
    G_{\rm SC}^{0\sigma_{2},L-1\sigma_{1}}(\vv{k},t_2,t_1),
\end{align}
\begin{align}
    &\vec{\chi}_{2}(t_3,t)
    \notag\\
    &= -i\Tr \int_{\mathcal{C}} dt_1
    G_{\rm NM}^{L\sigma_1,L\sigma'}(\vv{k},t_1,t)
    \vec{\tilde{\sigma}}^{\sigma'\sigma}
    G_{\rm SC}^{L-1\sigma,0\sigma_{3}}(\vv{k},t',t_3)
    \notag\\
    &\times
    \tau_e\sigma_{+}^{\sigma_{3},\sigma_{3}'}
    G_{\rm SC}^{0\sigma_{3}',L-1\sigma_{1}}(\vv{k},t_3,t_1),
\end{align}
Before expressing the component $\vec{J}_{\rm s, con}^{(3)}(L,t,t_1,t_2,t_3)$, we consider the pumped spin current defined as 
\begin{align}
    \Delta \vec{J}_{\rm s}^{(3)}(L,\omega_{\rm s}) := \vec{J}_{\rm s}^{(3)}(L,\omega_{\rm s}) - \into{\vec{J}_{\rm s}^{(3)}}{h=0}(L,\omega_{\rm s}),
\end{align}
where $\vec{J}_{\rm s}(L,\omega_{\rm s})$ is the Fourier component of the spin current in the presence of the external field.
This step is implemented by replacing $G_{\rm M}(\vv{q},\omega)$ with $\Delta G_{\rm M}(\vv{q},\omega)$ because the external field couples only with the magnon.

Here, in the Fourier component of terms of $\vec{\chi}_{1}(t,t_2)$ and $\vec{\chi}_{2}(t,t_2,t_3)$, we can find the delta function $\delta(\omega)$.
However, $\Delta G_{\rm M}(\vv{q},\omega)$ is include the delta function $\delta(\omega - \Omega)$.
Thus, the terms of $\vec{\chi}_{1}(t,t_2)$ and $\vec{\chi}_{2}(t,t_2,t_3)$ are vanished.
Therefore, the only non-vanishing terms are $\vec{J}_{\rm s, con}^{(3)}(L,t,t_1,t_2,t_3)$.
According to this discussion, the pumped spin current is written in the real-time representation as
\begin{align}
    &\Delta\vec{J}_{\rm s}^{(3)}(L,0)
    \notag\\
    &= -\frac{T_{\rm int}^2 V^{2}}{6} \lim_{t,t' \to 0}
    \Re
    \left[
    -i\Tr  
    \right.
    \notag\\
    &
    \vec{\sigma}\tau_z \star G_{\rm SC} \star \bar{\sigma}_{-} \Sigma_{3} \bar{\sigma}_{+}\star G_{\rm SC} \star G_{\rm NM}(t,t')
    \notag\\
    &+\left.\vec{\sigma}\tau_z \star G_{\rm SC} \star \bar{\sigma}_{+} \Sigma_{4}  
        \bar{\sigma}_{-} \star G_{\rm SC} \star G_{\rm NM}(t,t')
        \right],
        \label{pumped-spin-current-full}
\end{align}
where the "self-energy" are given as
\begin{align}
    \Sigma^{\sigma_2,\sigma_3}_{3}(\vv{k},t_2,t_3) &= G_{\rm SC}^{0\sigma_2,0\sigma_3}(\vv{k},t_2,t_3) \Delta G_{\rm M}(\vv{0},t_2,t_3),
    \\
    \Sigma^{\sigma_3',\sigma_2'}_{4}(\vv{k},t_3,t_2) &= G_{\rm SC}^{0\sigma_3',0\sigma_2'}(\vv{k},t_3,t_2) \Delta G_{\rm M}(\vv{0},t_2,t_3),
\end{align}
and each component of these "self-energy" are defined as
\begin{align}
    \Sigma^{<\sigma_2,\sigma_3}_{3}(\vv{k},t_2,t_3) &= G_{\rm SC}^{<,0\sigma_2,0\sigma_3}(\vv{k},t_2,t_3) \Delta G^{<}_{\rm M}(\vv{0},t_2,t_3)
    \notag\\
    \Sigma^{A\sigma_2,\sigma_3}_{3}(\vv{k},t_2,t_3) &= G_{\rm SC}^{A,0\sigma_2,0\sigma_3}(\vv{k},t_2,t_3) \Delta G^{<}_{\rm M}(\vv{0},t_2,t_3)
    \notag\\
    \Sigma^{R\sigma_2,\sigma_3}_{3}(\vv{k},t_2,t_3) &= G_{\rm SC}^{R,0\sigma_2,0\sigma_3}(\vv{k},t_2,t_3) \Delta G^{<}_{\rm M}(\vv{0},t_2,t_3).
\end{align}
The components of $\Sigma_{4}(\vv{k},t_3,t_2)$ are defined in the same way.

\begin{figure}
    \centering
    \includegraphics[scale=0.8]{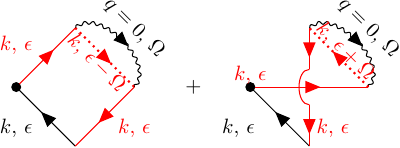}
    \caption{Feynman diagrams of the pumped spin current $\Delta J_{\rm s, con}^{(3)}(\vv{k},\epsilon,\omega)$.
    This is the Fourier component of the diagrams shown in FIG.~\ref{fig:time-dia}.
    }
    \label{fig:freq-dia}
\end{figure}
Consequently, the pumped spin current is written as
\begin{align}
    &\Delta \vec{J}_{\rm s}^{(3)}(L,\omega_{\rm s}) 
    \notag\\
    &= -\frac{T_{\rm int}^2 V^{2}}{6}\int d\epsilon d\omega \sum_{\vv{k} \in \rm BZ} \Delta J_{\rm s, con}^{(3)}(\vv{k},\epsilon,\omega)
\end{align}
where the $\Delta J_{\rm s, con}^{(3)}(\vv{k},\epsilon,\omega)$ is shown in FIG.~\ref{fig:freq-dia}.
More precisely, the pumped spin current is given by
\begin{widetext}
\begin{align}
    \Delta \vec{J}_{\rm s}^{(3)}(L,\omega_{\rm s}) 
    =& \frac{ \pi (\lambda h T_{\rm int} V)^{2}}{3} \abs{G_{\rm M}^{(0)\rm R}(\vv{0},\Omega)}^2 \delta(\omega_{\rm s}) \int d\epsilon \sum_{\vv{k} \in \rm BZ}d\vv{k} \Re \Tr 
    \notag\\ 
    & \vec{\tilde{\sigma}}
    \left[
        G_{\rm SC}^{R,L-1,0}(\vv{k},\epsilon)  \bar{\sigma}_{-} G_{\rm SC}^{R,0,0}(\vv{k},\epsilon-\Omega) \bar{\sigma}_{+} G_{\rm SC}^{R,0,L-1}(\vv{k},\epsilon) \tau_z  G_{\rm NM}^{<,L,L}(\vv{k},\epsilon) 
    \right.
    \notag\\
    &+ G_{\rm SC}^{R,L-1,0}(\vv{k},\epsilon)  \bar{\sigma}_{-} G_{\rm SC}^{R,0,0}(\vv{k},\epsilon-\Omega) \bar{\sigma}_{+} G_{\rm SC}^{<,0,L-1}(\vv{k},\epsilon)\tau_z G_{\rm NM}^{A,L,L}(\vv{k},\epsilon)
    \notag\\
    &+ G_{\rm SC}^{R,L-1,0}(\vv{k},\epsilon)  \bar{\sigma}_{-} G_{\rm SC}^{<,0,0}(\vv{k},\epsilon-\Omega) \bar{\sigma}_{+} G_{\rm SC}^{A,0,L-1}(\vv{k},\epsilon)\tau_z G_{\rm NM}^{A,L,L}(\vv{k},\epsilon)
    \notag\\
    &+ G_{\rm SC}^{<,L-1,0}(\vv{k},\epsilon)  \bar{\sigma}_{-} G_{\rm SC}^{A,0,0}(\vv{k},\epsilon-\Omega) \bar{\sigma}_{+} G_{\rm SC}^{A,0,L-1}(\vv{k},\epsilon) \tau_z G_{\rm NM}^{A,L,L}(\vv{k},\epsilon)
    \notag\\ 
    &+ G_{\rm SC}^{R,L-1,0}(\vv{k},\epsilon)  \bar{\sigma}_{+} G_{\rm SC}^{R,0,0}(\vv{k},\epsilon+\Omega) \bar{\sigma}_{-} G_{\rm SC}^{R,0,L-1}(\vv{k},\epsilon) \tau_z G_{\rm NM}^{<,L,L}(\vv{k},\epsilon)
    \notag\\
    &+ G_{\rm SC}^{R,L-1,0}(\vv{k},\epsilon)  \bar{\sigma}_{+} G_{\rm SC}^{R,0,0}(\vv{k},\epsilon+\Omega) \bar{\sigma}_{-} G_{\rm SC}^{<,0,L-1}(\vv{k},\epsilon) \tau_z G_{\rm NM}^{A,L,L}(\vv{k},\epsilon)
    \notag\\
    &+ G_{\rm SC}^{R,L-1,0}(\vv{k},\epsilon)  \bar{\sigma}_{+} G_{\rm SC}^{<,0,0}(\vv{k},\epsilon+\Omega) \bar{\sigma}_{-} G_{\rm SC}^{A,0,L-1}(\vv{k},\epsilon) \tau_z G_{\rm NM}^{A,L,L}(\vv{k},\epsilon)
    \notag\\
    &+\left. G_{\rm SC}^{<,L-1,0}(\vv{k},\epsilon)  \bar{\sigma}_{+} G_{\rm SC}^{A,0,0}(\vv{k},\epsilon+\Omega) \bar{\sigma}_{-} G_{\rm SC}^{A,0,L-1}(\vv{k},\epsilon) \tau_z G_{\rm NM}^{A,L,L}(\vv{k},\epsilon)
    \right],
    \label{app-pumped-spin-current}
\end{align}
\end{widetext}
where we used the relationships 
\begin{align}
    \Delta G^{<}(\vv{q},t_1,t_2) &=  \Delta G^{>}(\vv{q},t_1,t_2),
    \notag\\ 
    \LB \Delta G(\vv{q},t_1,t_2) \RB ^{*} &= -\Delta G(\vv{q},t_2,t_1).
\end{align}
Therefore, the pumped spin current is given by the static value $\omega_{\rm s} = 0$, and we reach to the formulation of DC spin current.
For the simplicity, we represent this value as 
\begin{align}
    \Delta \vec{J}_{\rm s}^{(3)}(L,\omega_{\rm s}) 
    &= J_{0} \delta(\omega_{s}) \vec{\mathcal{J}}_{\rm s}(L),
    \label{app-pumped-spin-current-simp}
    \\
    J_{0} &:= \frac{ \pi (\lambda h T_{\rm int} V)^{2}}{3} \abs{G_{\rm M}^{(0)\rm R}(\vv{0},\Omega)}^2.
\end{align}

\section{Green's function of the SC layer}\label{app-SC}
In this section, we show the method to calculate the real-space Green's function of the SC layer defined as
\begin{align}
    G_{\rm SC}(\vv{k},t_1,t_2) &:= -i \expval{T_{\mathcal{C}} d_{L-1/0\sigma_{1},\vv{k}}(t_1) d_{0/L-1\sigma',\vv{k}}^{\dag}(t) },
\end{align}
where the action used to derive this function is based on Hamiltonians.
\begin{align}
    H &= H_{\rm FI} + H_{\rm SC} + H_{\rm NM} + H_{\rm int}, 
    \\ 
    H_{\rm FI} &= -\sum_{i,i'\in \Lambda_{\rm FI}}\sum_{\sigma,\sigma'} \sum_{\vv{k}} \hat{c}_{i\sigma,\vv{k}}^{\dag} \mathcal{H}^{i\sigma,i'\sigma'}_{\rm FI}(\vv{k}) \hat{c}_{i'\sigma',\vv{k}}, 
    \\
    H_{\rm SC} &= \sum_{i,i'\in \Lambda_{\rm SC}}\sum_{\sigma,\sigma'}\sum_{\vv{k}} \hat{d}_{i\sigma,\vv{k}}^{\dag} \mathcal{H}^{i\sigma,i'\sigma'}_{\rm SC}(\vv{k}) \hat{d}_{i'\sigma',\vv{k}},
    \\
    H_{\rm NM} &= \sum_{i,i'\in \Lambda_{\rm NM}}\sum_{\sigma,\sigma'} \sum_{\vv{k}} \hat{c}_{i\sigma,\vv{k}}^{\dag} \mathcal{H}^{i\sigma,i'\sigma'}_{\rm NM}(\vv{k}) \hat{c}_{i'\sigma',\vv{k}},
    \\
    H_{\rm T} &=T_{\rm int}  \sum_{\sigma\vv{k}} 
        \LB
        \hat{d}_{L\sigma,\vv{k}}^{\dag} \tau_{z} \hat{c}_{L\sigma,\vv{k}}
        +\hat{d}_{0\sigma,\vv{k}}^{\dag} \tau_{z} \hat{c}_{0\sigma,\vv{k}}   +\text{h.c.}
        \RB,
    \\ 
    \mathcal{H}^{i\sigma,i'\sigma'}_{\rm FI}(\vv{k}) &= \delta_{i,i'}(-2t \cos k - M) \sigma_{z}^{\sigma\sigma'} - \delta_{i,i'\pm1}t\sigma_{z},
\end{align}
where $H_{\rm FI}$ is the Hamiltonian of the electron in the ferromagnetic insulator defined on the lattice $\Lambda_{\rm FI} = \lB i \mid i \leq 0,\, i \in \mathbb{Z} \rB$.
Here, the physical meaning of the $2n$th-order perturbation of the tunneling Hamiltonian is the same as the $n$th-order perturbation of the s-d interaction in $H_{\rm int}$.
Therefore, this system is equivalent to the system in the main text.
According to the second-order perturbation expansion of the tunneling Hamiltonian, the lesser part of the Green's function of the SC layer is written as
\begin{align}
    &G_{\rm SC}^{i\sigma,j\sigma'(<)}(\vv{k},\epsilon)
    \notag\\
    &= G_{\rm SC}^{i\sigma,m\sigma_1(R)}(\vv{k},\epsilon) \Sigma_{\rm T}^{m\sigma_1,n\sigma_2(<)}(\vv{k},\epsilon) G_{\rm SC}^{n\sigma_2,j\sigma'(A)}(\vv{k},\epsilon),
\end{align}
where $\Sigma_{\rm T}^{m\sigma_1,n\sigma_2(<)}(\vv{k},\epsilon)$ is the lesser part of the self-energy on the surface defined as 
\begin{align}
    \Sigma_{\rm T}^{0\sigma_1,0\sigma_2(R)}(\vv{k},\epsilon)
    &= T_{\rm int}^2 \tau_z G_{\rm FI}^{-1\sigma_1,-1\sigma_2(R)}(\vv{k},\epsilon) \tau_z ,
    \notag\\
    \Sigma_{\rm T}^{L\sigma_1,L\sigma_2(R)}(\vv{k},\epsilon)
    &= T_{\rm int}^2 \tau_z G_{\rm NM}^{L\sigma_1,L\sigma_2(R)}(\vv{k},\epsilon) \tau_z ,
    \notag\\ 
    \Sigma_{\rm T}^{i\sigma_1,j\sigma_2(R)}(\vv{k},\epsilon) &= 0 \quad \text{otherwise}.
\end{align}
Using this expression, the lesser part is written as 
\begin{align}
    &\Sigma_{\rm T}^{m\sigma_1,n\sigma_2(<)}(\vv{k},\epsilon) 
    = 2i F^{m,n}(\epsilon) \Im \Sigma_{\rm T}^{m\sigma_1,n\sigma_2(R)}(\vv{k},\epsilon),
    \notag\\ 
    &F^{0,0}(\epsilon)
    = \mqty(
        f(\epsilon - \Omega) & 0  & 0 & 0 \\ 
        0 & f(\epsilon + \Omega) & 0 & 0 \\
        0 & 0 & f(\epsilon + \Omega) & 0 \\
        0 & 0 & 0 & f(\epsilon - \Omega)
    ),
    \notag\\
    &F^{L,L}(\epsilon)
    = \mqty(
        f(\epsilon) & 0  & 0 & 0 \\ 
        0 & f(\epsilon) & 0 & 0 \\
        0 & 0 & f(\epsilon) & 0 \\
        0 & 0 & 0 & f(\epsilon)
    ),
\end{align}
where $f(\epsilon) = 1/(e^{\beta \epsilon} + 1)$.
The retarded and advanced part of the Green's function is part of the contour Green's function defined as
\begin{align}
    G_{\rm SC}(\vv{k},t_1,t_2) &:= -i \expval{T_{\mathcal{C}} d_{L-1/0\sigma_{1},\vv{k}}(t_1) d_{0/L-1\sigma',\vv{k}}^{\dag}(t) }_0,
\end{align}
where this expectation value is based on the theory without interactions.
Therefore, the retarded Green's function is defined as 
\begin{align}
    G_{\rm SC}^{i\sigma,j\sigma' (R)}(\vv{k},\epsilon)
    := \LB \lb \epsilon - \mathcal{H}_{\rm SC}(\vv{k}) + i\delta \rb^{-1} \RB^{i\sigma,j\sigma'}.
\end{align}
The advanced part is the counterpart of $i \delta \to -i\delta$.
To calculate the lesser part of the Green's function, we can use the knitting algorithm shown in Ref.~\onlinecite{Knit-Green}.

\section{Green's function of the NM layer}\label{app-NM}
In this section, we derive the surface Green's function of NM using the recursive Green's function.
The detail of this method is seen in Ref.~\onlinecite{Recursive-Takagi, Recursive-Umerski}.
As the model of NM, we consider the tight-binding Hamiltonian with spin-independent nearest-neighbor hopping (i.e., without spin-orbital coupling) defined as 
\begin{align}
    \left.\mathcal{H}_{\rm NM}^{i\sigma,i'\sigma'}\right|_{i\neq i'}
    &:= t(i,i') \delta_{\sigma,\sigma'} (\delta_{i,i'+1} + \delta_{i,i'-1}).
\end{align}
To treat the coupling between the SC, we must define the Hamiltonian on a Nambu basis.
The function $t(i,i') \in \mathbb{R}$ is a hopping parameter between the sites $i$ and $i'$.
For convenience, we assume that the sites are aligned linearly in a certain direction.
We perform the Fourier transformation in the plane perpendicular to the direction of $i$.
After this Fourier transformation, we can get the Hamiltonian 
\begin{align}
    \mathcal{H}_{\rm NM}^{i\sigma,i'\sigma'}(\vv{k})
    &= \delta_{i,i'}\hat{\mathcal{H}}^{\sigma\sigma'}_{0}(\vv{k}) + \delta_{i',i+1}\mathcal{T}_{\rm NM}+ \delta_{i',i-1}\mathcal{T}_{\rm NM}^{\dag}.
\end{align}
According to this Hamiltonian, the surface Green's function of the NM layer can be derived using the recursive Green's function method.
Therefore, the Green's function of the NM layer is written as
\begin{align}
    \hat{g}_{\rm NM}^{\infty}(\vv{k},\tilde{z})
    &= \hat{Q}_{12}\hat{Q}_{22}^{-1},
    \notag \\
    Q &:=
    \mqty(
        \hat{Q}_{11} & \hat{Q}_{12} \\
        \hat{Q}_{21} & \hat{Q}_{22}
    ),
    \notag \\
    Q^{-1}XQ &= \mathrm{spec}\lB\lambda_{i}\rB,
    \notag \\ 
    X &= \mqty(
        \hat{O} & \lb\hat{\mathcal{T}}_{\rm NM}^{\dag}\rb^{-1} \\
        -\hat{\mathcal{T}}_{\rm NM}  & (\tilde{z} + \hat{\mathcal{H}_{0}}(\vv{k})) \lb\hat{\mathcal{T}}_{\rm NM}^{\dag}\rb^{-1}
    ),
\end{align}
where $\hat{O}$ and $\hat{I}$ are the $4\times4$ zero and identity matrix, respectively.
$\lambda_{i} \in \mathbb{C}$ is the eigenvalue of the matrix $X$ with $|\lambda_{i}| \leq |\lambda_{i+1}|$.
Finally, $\tilde{z} = \omega -i\delta$ for advanced Green's function, and
$\tilde{z} = \omega + i\delta$ for retarded Green's function.
Here, we define the non-perturbed Green's function of NM as 
\begin{align}
    G_{\rm NM}^{\mathrm{R},L\sigma,L\sigma'(0)}(\vv{k},\epsilon) := \lb \hat{g}_{\rm NM}^{\infty}\rb^{\sigma\sigma'} (\vv{k},\epsilon+i\delta).
\end{align}

\bibliography{ref}

\end{document}